\documentclass[11pt,letterpaper]{article}
\usepackage{orcidlink}
\usepackage[T1]{fontenc}
\usepackage[utf8]{inputenc}
\usepackage[margin=1in]{geometry}
\usepackage{amsmath,amsfonts}
\IfFileExists{newtxtext.sty}{%
  \usepackage{newtxtext,newtxmath}%
}{%
  \usepackage{mathptmx}
}
\usepackage{microtype}

\usepackage{graphicx}
\graphicspath{{./}{figures/}}
\usepackage{booktabs,array,tabularx}
\usepackage{enumitem}
\usepackage{xcolor}
\usepackage{caption}
\usepackage[noblocks]{authblk}

\usepackage[title]{appendix}
\numberwithin{equation}{section}

\usepackage[numbers,square,sort&compress]{natbib}
\usepackage{hyperref}
\hypersetup{
  colorlinks=true,
  allcolors=blue,
  plainpages=false,
  breaklinks=true,
  pdfborder={0 0 0},
  pdftitle={Diff-NekRS: A Scalable Differentiable Framework for Multi-Timestep Solver-in-the-Loop Training},
  pdfauthor={Junoh Jung, Riccardo Balin, Bethany Lusch, Emil Constantinescu}
}

\renewenvironment{abstract}
  {\begin{quote}
   \noindent\rule{\linewidth}{0.5pt}\par
   \noindent\textbf{\abstractname.}\par\noindent\ignorespaces}
  {\par\medskip\noindent\rule{\linewidth}{0.5pt}
   \end{quote}}
\newenvironment{keywords}
  {\begin{quote}\small\noindent\textbf{Keywords: }\ignorespaces}
  {\par\end{quote}}

\newcommand{\nekrs}{NekRS}

\newcommand{\libtorch}{LibTorch}
\newcommand{\aurora}{Aurora}
\newcommand{\firedrake}{Firedrake}
\providecommand{\Description}[1]{}

\title{\bfseries Diff-NekRS: A Scalable Differentiable Framework for
Multi-Timestep Solver-in-the-Loop Training}
\author[1]{Junoh Jung\thanks{Corresponding author:
\href{mailto:jjung@anl.gov}{\texttt{jjung@anl.gov}}.}}
\author[2]{Riccardo Balin}
\author[2]{Bethany Lusch}
\author[1]{Emil Constantinescu}
\affil[1]{Mathematics and Computer Science Division, Argonne National Laboratory,
Lemont, IL 60439, USA}
\affil[2]{Argonne Leadership Computing Facility, Argonne National Laboratory,
Lemont, IL 60439, USA}
\date{}

\begin{document}
\maketitle
\vspace{-1cm}

\begin{abstract}
Hybrid physics--machine-learning solvers improve under-resolved simulations by embedding trainable corrections into the time integration. During autoregressive inference, repeated solver--model interactions can amplify small errors, motivating multi-timestep solver-in-the-loop training. However, production solvers rarely expose the derivatives needed to backpropagate through such rollouts. We introduce Diff-NekRS, a scalable differentiable framework that embeds neural corrections directly in the GPU-accelerated NekRS incompressible-flow solver. NekRS computes the authoritative forward trajectory, a manually implemented exact discrete adjoint differentiates the supported fully discrete timestep, and LibTorch supplies neural vector--Jacobian products and parameter gradients. End-to-end Taylor and centered finite-difference tests verify the assembled gradient for two-dimensional cylinder flow (2Dcyl) and the three-dimensional Taylor--Green vortex (3DTGV) across five horizons and 12--1,020 MPI ranks. At 1,020 ranks, optimizer-enabled post-setup training updates retain 54.5\%--78.0\% and 80.7\%--81.9\% weak-scaling efficiency for 2Dcyl and 3DTGV, respectively. In 200-step autoregressive inference, the $M=50$ model reduces the three-seed median terminal relative $L_2$ velocity error by 59.2\% for 2Dcyl and 12.1\% for 3DTGV relative to the uncorrected coarse-grid $P=2$ baseline, and retains wall-clock speedups of $5.38\times$ and $2.49\times$, respectively, relative to the corresponding $P=7$ configurations for equal simulated-time intervals. These results establish a verified and scalable path for multi-timestep solver-in-the-loop training that improves coarse-grid trajectory accuracy while retaining a speed advantage over the high-order reference.

\end{abstract}

\begin{keywords}
scientific machine learning, solver-in-the-loop training, NekRS, LibTorch,
discrete adjoint, weak scaling, high-performance computing
\end{keywords}

\begin{figure}[!t]
    \centering
    \includegraphics[
        width=\textwidth,
        height=0.72\textheight,
        keepaspectratio
    ]{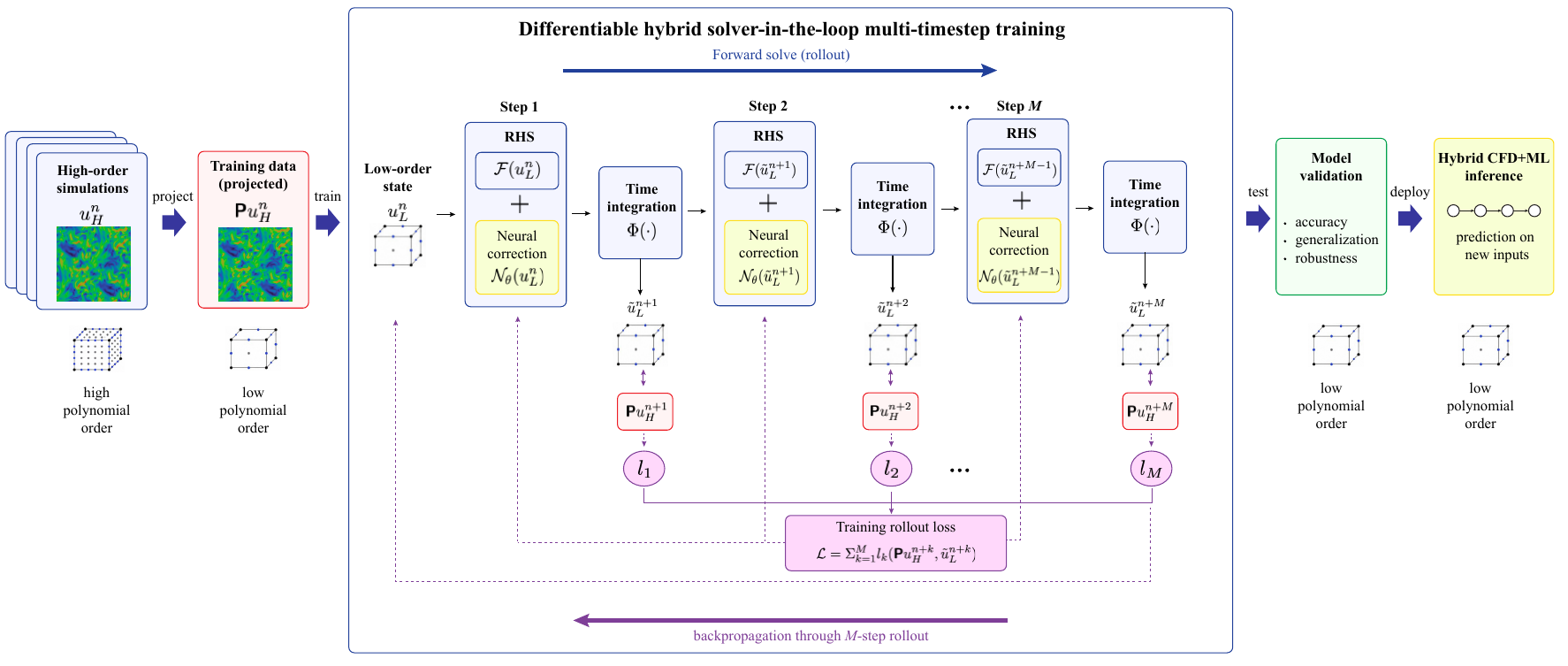}
    \caption{Multi-timestep solver-in-the-loop training in Diff-NekRS.}
    \label{fig:workflow}
\end{figure}

\section{Introduction}
Hybrid physics--machine learning (ML) models augment established numerical solvers with trainable closures, source terms, or discretization corrections that approximate unresolved dynamics or model-form error. In under-resolved computational fluid dynamics (CFD), this strategy can improve coarse-grid predictions without replacing the governing-equation solve or the core numerical machinery of the host solver~\cite{macart2021embedded,kochkov2021mlcfd}. One common approach is to evaluate the neural component within the time-advancement loop, so that its output influences the state passed to subsequent solver and model evaluations.

Many such models are nevertheless trained with one-step targets formed from paired under-resolved and high-fidelity states. In training, the neural model is evaluated on states extracted from a prescribed data trajectory, whereas at inference it receives states produced by preceding corrected steps. This mismatch creates a rollout distribution shift, such that small errors can alter later model inputs, accumulate along the trajectory, and degrade long-horizon accuracy or stability. Solver-in-the-loop training reduces this mismatch by unrolling the corrected solver for multiple timesteps and optimizing a trajectory-level objective. The learned component is thereby trained under the autoregressive conditions in which it will be used \cite{um2020solverintheloop,list2022learnedturbulence}.

A central challenge in this setting is gradient computation for multi-timestep objectives, which requires propagating sensitivities through the full coupled solver-model trajectory. For large-scale CFD solvers, implementing this reverse pass efficiently and correctly is difficult because the forward timestepper contains many interacting, state-dependent operations distributed across accelerators and MPI ranks. These requirements make scalable exact-gradient training in production solvers nontrivial.

We present Diff-NekRS, a NekRS-based framework for scalable differentiable multi-timestep solver-in-the-loop training that embeds a neural model in the GPU-accelerated NekRS incompressible-flow solver~\cite{fischer2022nekrs}. This work extends our earlier hybrid NekRS formulation \cite{jung2026hybrid} in two principal ways. First, it reformulates the learned correction from a post-step velocity increment as a neural momentum source integrated into the native timestep. Second, for the supported timestep configuration, it replaces the surrogate-gradient path with an exact discrete adjoint. Although the implemented solver-step vector--Jacobian product (VJP) is specific to the supported \nekrs{} timestep, the separation of the solver-step VJP, neural-source VJP, and distributed gradient assembly provides a reusable integration pattern for other production finite- and spectral-element solvers. We therefore evaluate the assembled gradient end-to-end across distributed MPI allocations and characterize the weak scaling of optimizer-enabled multi-horizon training updates on \aurora{} through 1,020 MPI ranks. We assess numerical correctness, training behavior, autoregressive accuracy, inference cost, and weak-scaling performance in two examples: two-dimensional flow past a cylinder (2Dcyl) and the three-dimensional Taylor--Green vortex (3DTGV).

The main contributions of this work are the following:
\begin{itemize}[
  leftmargin=*,
  noitemsep,
  topsep=2pt
]

  \item We develop an exact discrete adjoint for the supported \nekrs{} timestep configuration and couple it to \libtorch{} neural-model VJPs for multi-timestep training.

  \item We verify the full solver--model parameter gradient for two examples (2Dcyl and 3DTGV) at rollout horizons $M\in\{1,5,10,20,50\}$ across MPI allocations $N_r\in\{12,24,48,96,252,516,1020\}$.

  \item We evaluate multi-horizon exact-adjoint training with fixed held-out validation, quantify 200-step autoregressive accuracy relative to the uncorrected $P=2$ baseline, measure wall-clock speedups relative to the corresponding $P=7$ configurations, and characterize optimizer-enabled $B=4$ training updates on Aurora through 1,020 MPI ranks.

\end{itemize}

Fig.~\ref{fig:workflow} illustrates the multi-timestep training process within a hybrid physics--ML solver. High-order trajectories are first projected onto the coarse grid to generate training targets. The \nekrs{} solver then advances the state using neural source corrections, while the exact discrete adjoint propagates the trajectory-level loss backward through the \libtorch{} model. During training, the model is also validated before being deployed for inference.

\section{Related Work}
\label{sec:related-work}

Differentiable scientific    computing frameworks differentiate through numerical time integration or embed learned terms in partially known dynamics \cite{chen2018neuralode,rackauckas2019diffeqflux, rackauckas2020universal,hu2020difftaichi}. Automatic-differentiation-native CFD solvers have also been developed across several software ecosystems. For example, JAX-Fluids 2.0~\cite{bezgin2025jaxfluids2} supports differentiable, accelerator-based compressible flow simulation, PICT~\cite{franz2026pict} provides a GPU-accelerated differentiable multiblock PISO solver in PyTorch, and WaterLily.jl~\cite{weymouth2025waterlily} offers a backend-agnostic differentiable Julia solver for incompressible immersed-boundary flows.

In CFD, data-driven corrections include Reynolds stress discrepancies, subgrid forcing, and closures evaluated both a priori and a posteriori \cite{wang2017reynoldsdiscrepancy,maulik2019subgrid, macart2021embedded}. Solver-in-the-loop methods further optimize the learned term through repeated interaction with the numerical solver, reducing rollout-distribution mismatch and improving a posteriori behavior \cite{um2020solverintheloop,list2022learnedturbulence, fan2025neuraldifferentiable, shankar2025differentiableturbulence, fan2026coupledclosures}.

Our previous work on hybrid physics--ML methods spans differentiable discontinuous-Galerkin source corrections~\cite{kang2023subgridnode,kang2023differentiabledg}, weak-form correction~\cite{jung2026weakform}, and \nekrs{}-based models, including offline and surrogate-gradient multistep training \cite{jung2025hybrid,jung2026hybrid}. The \firedrake{} framework provides related continuous Galerkin and automatic-differentiation infrastructure through pyadjoint/dolfin-adjoint~\cite{rathgeber2017firedrake,Mitusch2019DolfinAdjoint}. In contrast to automatic-differentiation-native solver stacks, Diff-NekRS retains the established distributed C++/OCCA \nekrs{} forward solver and retrofits a manually implemented discrete timestep VJP for the supported time-advancement configuration. The solver VJP is coupled to \libtorch{} model VJPs and verified across rollout horizons and MPI allocations before evaluating training, inference, and weak scaling.

\section{Methods}
\label{sec:methods}
This section presents the hybrid physics--machine learning formulation, along with the differentiable learning strategy and implementation for the multi-timestep solver-in-the-loop training workflow. We first define the neural momentum-source coupling and the rollout objective. We then derive the discrete reverse recurrence for the implemented \nekrs{} timestep and describe the distributed \nekrs{}--\libtorch{} execution, including taped forward states, model vector--Jacobian products, global loss normalization, MPI gradient synchronization, and synchronized optimizer updates.

\subsection{Hybrid Physics--Machine Learning Model}
\label{sec:hybrid-model}
\begin{figure}[!t]
    \centering
    \includegraphics[
        width=\textwidth,
        height=0.72\textheight,
        keepaspectratio
    ]{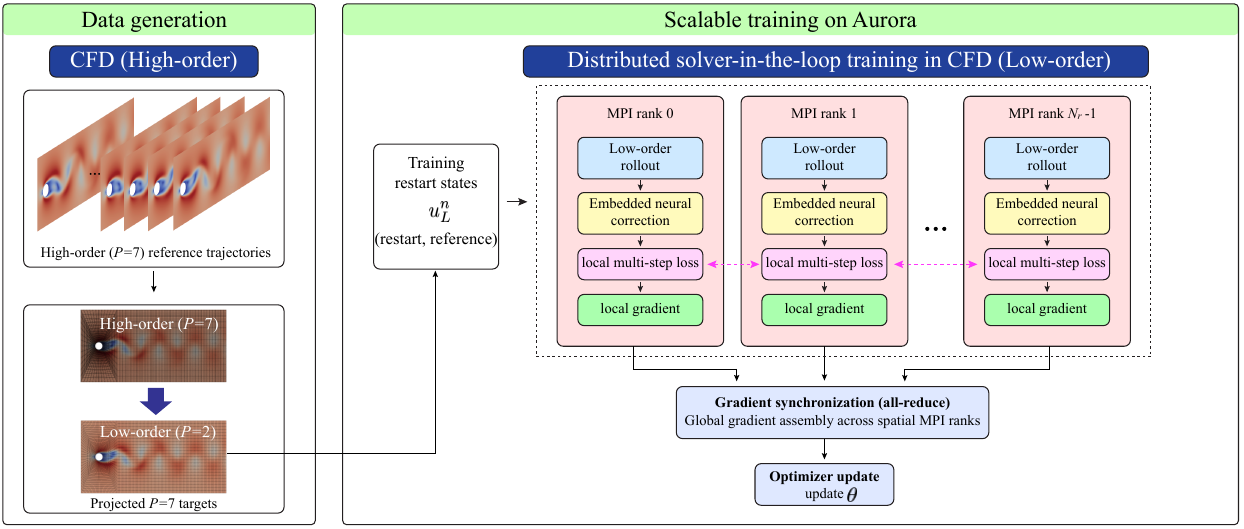}
\caption{Distributed solver-in-the-loop training workflow on
\aurora{}.}
\label{fig:distributed-workflow}
    \Description{One sampled rollout is spatially decomposed across MPI ranks.
    Each rank advances its local subdomain and forms a local gradient
    contribution. An all-reduce sums those contributions, every rank applies
    the identical Adam update, and rank zero writes the checkpoint.}
\end{figure}

We start with the spatially coarse-grid discretization of the incompressible Navier--Stokes operator, $\mathcal{F}_L$,

\begin{equation}
    \frac{\partial u_L}{\partial t}
    =
    \mathcal{F}_L
    \left(
        u_L,p_L;\xi
    \right),
    \label{eq:coarse-continuous}
\end{equation}
where $u_L(t)$ and $p_L(t)$ denote the low-fidelity coarse-grid velocity and pressure fields, respectively, and $\xi$ represents non-trainable solver parameters such as viscosity, boundary conditions, and discretization settings. We augment the coarse-grid model with a neural source term,
\begin{equation}
    \frac{\partial u_L}{\partial t}
    =
    \mathcal{F}_L
    \left(
        u_L,p_L;\xi
    \right)
    +
    s_{\theta}
    \left(
        u_L
    \right),
    \label{eq:hybrid-source-abstract}
\end{equation}
where $s_{\theta}$ is a trainable, source-like correction.  In strong form, the corrected incompressible momentum equations are
\begin{align}
    \frac{\partial u_L}{\partial t}
    +
    \left(
        u_L \cdot \nabla
    \right)u_L
    &=
    -\nabla p_L
    +
    \nu \nabla^2 u_L
    +
    f_L
    +
    s_{\theta}
    \left(
        u_L
    \right),
    \label{eq:hybrid-source-momentum}
    \\
    \nabla \cdot u_L &= 0,
    \label{eq:hybrid-source-continuity}
\end{align}
where $f_L$ denotes any prescribed physical forcing.  Thus, the neural model does not directly overwrite the velocity solution.  Instead, it predicts an additional forcing or unresolved-physics term that is integrated by the native \nekrs{} timestepper.
For each spectral element $e$, a local stencil operator $S_e$ gathers the velocity field from the target element and its face-adjacent neighbors. At timestep $n$, each MPI rank constructs the neural input for every locally owned element $e$ from its element-centered velocity stencil:
\begin{align}
    q_e^n
    &=
    S_e\!\left(
        u_L^n
    \right),
    \label{eq:source-network-input}
    \\
    \widetilde q_e^n
    &=
    \mathcal{N}\!\left(q_e^n\right),
    \label{eq:normalized-source-input}
    \\
    \widehat{s}_{\theta,e}^n
    &=
    \tanh\!\left(
        N_{\theta}\!\left(\widetilde q_e^n\right)
    \right),
    \label{eq:local-neural-source}
\end{align}
where $S_e$ gathers the velocity values from element $e$ and its face-adjacent neighbors and $\mathcal{N}$ denotes the
componentwise input normalization, followed by clipping to $[-10,10]$. Neighbor slots that fall outside the physical domain are set to zero after normalization, so the padding corresponds to zero in the normalized coordinates. The \libtorch{} model $N_{\theta}$ is a multilayer perceptron with two 256-unit GELU hidden layers and a linear output layer. The subsequent elementwise $\tanh$ activation bounds each dimensionless output component to $[-1,1]$. Accordingly, $\widehat{s}_{\theta,e}^n$ is the bounded, element-local neural source prediction, and it is subsequently projected to have zero global componentwise mean. The resulting element-local predictions are inserted into the native spectral-element momentum right-hand side,
\begin{equation}
    s_{\theta}^n
    =
     B_{\mathrm{RHS}}
    \left(
        \Pi_0
        \left\{
            \widehat s_{\theta,e}^n
        \right\}_{e\in\mathcal{E}}
    \right),
    \label{eq:assembled-neural-source}
\end{equation}
where $\Pi_0$ is the distributed componentwise zero-mean projection and
$B_{\mathrm{RHS}}$ denotes insertion into the native \nekrs{} element-local
right-hand-side storage together with the solver's standard assembly and
boundary-constraint operations.

Unlike the post-step velocity corrections used previously~\cite{jung2025hybrid,jung2026hybrid}, the transformed predictions $\widehat{s}_{\theta,e}^{n}$ are zero-mean projected and assembled according to Eq.~\eqref{eq:assembled-neural-source} to form $s_{\theta}^{n}$, which enters the momentum right-hand side and is integrated by the solver. The per-step influence of $s_{\theta}^{n}$ is therefore mediated by the discrete time-advancement and pressure-projection operators and typically scales with $\Delta t$. The source-based formulation can thus produce smaller immediate changes in the velocity field, leading to more gradual one-step loss reduction than with a direct velocity
update.  

The resulting fully discrete solver--model step is
\begin{equation}
    u_L^{n+1}
    =
    F_n(u_L^n,\theta)
    \equiv
    \Phi_{\Delta t_n}^{h}
    \left(
        u_L^n;
        s_{\theta}^n
    \right),
    \qquad
    s_{\theta}^n=s_{\theta}(u_L^n).
    \label{eq:exact-discrete-map}
\end{equation}
Here, $s_{\theta}^n$ denotes the assembled neural right-hand-side contribution defined in Eq.~\eqref{eq:assembled-neural-source}. For a given run, $h$ collectively denotes the spatial discretization, including the element mesh, polynomial order, quadrature, and discrete boundary treatment. The map $\Phi_{\Delta t_n}^{h}(u;s)$ denotes one \nekrs{} timestep with the prescribed assembled source contribution $s$, and it includes the remaining time-advancement operations, pressure projection, velocity boundary enforcement, and the algebraic elliptic solution operators at the configured tolerances. The composite map $F_n$ additionally includes evaluation and assembly of the neural source from $u_L^n$. The implementation does not invoke a separate learned-correction shared-node averaging operator.

For a rollout beginning at timestep $k$, the composite
solver--model step is applied autoregressively for $M$ timesteps,
\begin{equation}
\begin{aligned}
    u_L^{k+j}
    =
    \Phi_{\Delta t_{k+j-1}}^{h}
    \left(
        u_L^{k+j-1};
        s_{\theta}^{k+j-1}
    \right),
    \qquad j=1,\ldots,M.
\end{aligned}
\label{eq:source-rollout}
\end{equation}

Let $\mathcal{P}$ denote the projection operator from the high-fidelity space onto the coarse-grid space. The corresponding coarse-grid reference state at timestep $k+j$ is defined as
\begin{equation}
    u_{\mathrm{ref}}^{k+j}
    =
    \mathcal{P}u_H^{k+j},
\end{equation} 
where $u_H(t)$ denotes the high-fidelity velocity field on the fine grid. The multi-timestep training objective is
\begin{equation}
    \mathcal{L}_M(\theta)
    =
    \frac{1}{2}
    \sum_{j=1}^{M}
    \frac{w_j}{\mathcal{N}_{\omega}}
    \left\|
      D_{\omega}\!\left(u_L^{k+j}-u_{\mathrm{ref}}^{k+j}\right)
    \right\|_{M_L}^{2},
    \label{eq:source-rollout-loss}
\end{equation}
where $w_j$ denotes the weight assigned to rollout step $j$ and $\|v\|_{M_L}^{2}=v^T M_Lv$, where $M_L$ is the coarse-grid mass matrix. The fixed binary diagonal mask \(D_\omega\) selects the spatial region used for training, and
$\mathcal{N}_{\omega}$ is the globally reduced sum of the corresponding masked mass weights.  This normalization makes the objective a distributed, mass-weighted squared-error objective. The parameter gradient can therefore be written as
\begin{equation}
    \frac{d\mathcal{L}_M}{d\theta}
    =
    \sum_{n=k}^{k+M-1}
    \left(
        \frac{\partial s_{\theta}^n}{\partial\theta}
    \right)^{\!\top}
    g_s^n,
    \qquad
    g_s^n
    =
    \frac{\partial\mathcal{L}_M}
         {\partial s_{\theta}^n}.
    \label{eq:source-gradient-interface}
\end{equation}
For the production runs in this paper, $g_s^n$ is supplied by the exact
discrete adjoint of the supported \nekrs{} timestep.

\subsection{Differentiable Learning}
\label{sec:differentiable-learning}

Training requires the gradient of the multi-timestep rollout loss with respect to the neural-network parameters. A manually implemented algebraic discrete adjoint of the fully discrete NekRS timestep computes the loss sensitivity with respect to the source ($g^n_s$) at each rollout step, and LibTorch applies the chain rule through the neural model---the interface between NekRS and the learning system---to obtain the parameter gradient. The coupled forward and reverse dependencies are

\begin{equation}
\begin{aligned}
\text{Forward:}\quad
& \theta
  \longrightarrow s_{\theta}^{n}
  \longrightarrow u_L^{n+1}
  \longrightarrow \mathcal{L}_M,
\\
\text{Reverse:}\quad
& \mathcal{L}_M
  \xrightarrow{\text{\nekrs{} adjoint}}
  g_s^n,
\\[-1mm]
& g_s^n
  \xrightarrow{\text{\libtorch{} VJP}}
  \nabla_{\theta}\mathcal{L}_M.
\end{aligned}
\label{eq:solver-model-gradient-flow}
\end{equation}

We now describe the discrete adjoint of the composite solver--model step $F_n$ defined in Eq.~\eqref{eq:exact-discrete-map}. Writing the rollout objective as $\mathcal{L}_M=\sum_{j=1}^{M} \ell_{k+j}(u_L^{k+j})$ and defining $\ell_k\equiv0$, reverse accumulation starts with

\begin{equation}
    \lambda^{k+M}
    =
    \nabla_{u}\ell_{k+M}(u_L^{k+M})
    \label{eq:exact-terminal-adjoint}
\end{equation}
and proceeds backward for $n=k+M-1,\ldots,k$ according to
\begin{equation}
    \lambda^n
    =
    \left(\frac{\partial F_n}{\partial u_L^n}\right)^{\!T}
    \lambda^{n+1}
    +
    \nabla_u\ell_n(u_L^n).
    \label{eq:exact-discrete-recurrence}
\end{equation}
The source and parameter gradients are
\begin{equation}
    \begin{aligned}
        g_s^n
        &=
        \left(\frac{\partial \Phi_{\Delta t_n}^{h}}
        {\partial s_{\theta}^n}\right)^{\!T}\lambda^{n+1},
        \\
        \nabla_{\theta}\mathcal{L}_M
        &=
        \sum_{n=k}^{k+M-1}
        \left(\frac{\partial s_{\theta}^n}{\partial\theta}\right)^{\!T}
        g_s^n.
    \end{aligned}
    \label{eq:exact-source-parameter-gradient}
\end{equation}
The state contribution $(\partial s_{\theta}^n/\partial u_L^n)^Tg_s^n$ is included in Eq.~\eqref{eq:exact-discrete-recurrence}. The \nekrs{}--\libtorch{} interface applies the transpose actions associated with source assembly, the zero-mean projection, normalization, clipping, and stencil/halo construction, while \libtorch{} computes the model-input and parameter VJPs through
$ \tanh\!\left(N_\theta(\widetilde q_e^n)\right)$.

Although the reverse recurrence applies the linearization of the nonlinear composite timestep, this does not make the resulting gradient approximate. Every first-order adjoint of a nonlinear map applies a Jacobian transpose evaluated along the nonlinear primal trajectory. Here, the recurrence is linear in $\lambda^n$, while its coefficients depend on $u_L^n$ \cite{giles2000adjoint,farrell2013automated}.  Reverse-mode automatic differentiation of the same nonlinear timestep would construct this same linear transpose-Jacobian recurrence rather than a nonlinear equation in the adjoint variable.
\subsection{Implementation of Distributed \nekrs{}--\libtorch{} Training}

This section describes the end-to-end implementation of the distributed solver-in-the-loop training workflow. As summarized
in Fig.~\ref{fig:distributed-workflow}, the workflow consists of four stages: fine-grid reference data generation and projection, source-corrected coarse-grid rollouts, reverse differentiation, and synchronized model updates. High-order $P=7$ reference trajectories are first generated using NekRS and projected onto the $P=2$ space. The projected fields provide timestep-aligned training targets and restart states for the coarse-grid rollouts.

\begin{table}[t]
\caption{Responsibilities in the distributed
forward-reverse training path.}
\label{tab:training-routines}
\centering
\small
\setlength{\tabcolsep}{4pt}
\renewcommand{\arraystretch}{1.12}
\begin{tabularx}{\linewidth}{@{}
  >{\raggedright\arraybackslash}p{0.23\linewidth}
  >{\raggedright\arraybackslash}p{0.25\linewidth}
  >{\raggedright\arraybackslash}X@{}}
\toprule
Stage & Implementation & Responsibility \\
\midrule

\textbf{Forward rollout}
&
\nekrs{} + \libtorch{}
&
Evaluate the neural source, advance the authoritative state,
accumulate the loss, and retain $u_L^n$ and $\Delta t_n$.
\\

\textbf{Reverse sweep}
&
\nekrs{} adjoint + \libtorch{}
&
Propagate the state adjoint, form the neural-source seed, and
accumulate model-input and parameter VJPs.
\\

\textbf{Global update}
&
MPI + Adam
&
Sum rank-local parameter-gradient contributions, clip the resulting global
gradient, apply the identical Adam update on every rank, and checkpoint on
rank~0.
\\

\bottomrule
\end{tabularx}

\end{table}

Table~\ref{tab:training-routines} summarizes the division of responsibility among \nekrs{}, \libtorch{}, and MPI. The forward rollout is executed inside the \nekrs{} timestep loop. At timestep $n$, the \libtorch{} model evaluates $\widehat{s}_\theta(u_L^n)$ from a normalized face-neighbor stencil, and \nekrs{} advances the authoritative
coarse-grid solution using the supported NekRS timestep. The corresponding loss contribution is accumulated, and the
resulting forward state $u_L^{n+1}$ and the actual timestep size $\Delta t_n$ used for the update are recorded for the reverse sweep. The corrected state $u_L^{n+1}$ is then used as the input to the next solver and neural-model evaluations. The training rollout therefore follows the same autoregressive coupling used during inference.

After the $M$-step forward rollout, an explicit reverse sweep traverses the recorded timestep intervals in reverse order. The
complete \nekrs{} trajectory is not represented by a global \libtorch{} computation graph, meaning that the backward pass 
does not differentiate the solver through a single \texttt{loss.backward()} call. At each reverse step, the exact
discrete solver-step vector--Jacobian product propagates the state adjoint and produces the neural-source sensitivity $g_s^n
=\frac{\partial \mathcal{L}_M}{\partial s_\theta^n}$. This sensitivity is passed to the \libtorch{} model VJP, which
accumulates the corresponding parameter gradient contribution and returns the model input contribution to the adjoint of
$u_L^n$. Direct loss-gradient seeds are added at every loss-bearing rollout state. Consequently, the final parameter
gradient includes both the direct contribution of every one of the $M$ loss terms and their indirect dependence through all
subsequent corrected timesteps.

For each sampled trajectory--start pair, all MPI ranks jointly advance one domain-decomposed rollout with identical model parameters. The ranks are therefore spatially parallel rather than independent data-parallel replicas. MPI reductions assemble the globally normalized, masked mass-weighted objective and the
global parameter gradient for each rollout. Each optimizer update averages the MPI-assembled gradients from $B=4$ sequential, domain-parallel rollouts, where $B$ denotes the batch size, clips the result to unit norm, and applies one synchronized Adam update on every rank.

\section{Results}
\label{sec:results}

We evaluate Diff-NekRS on two canonical incompressible-flow benchmarks: a nominally two-dimensional circular-cylinder wake (2Dcyl) at $Re_D=100$, in the periodic vortex-shedding regime below the first three-dimensional wake instability~\cite{posdziech2007cylinder,barkley1996floquet}, and the triply periodic three-dimensional Taylor--Green vortex (3DTGV) at $Re=1600$, which represents vortex stretching, transition to turbulence, and decay~\cite{brachet1983tgv,vanrees2011tgv}. For 3DTGV training, coordinate-wise phase shifts generate ten training initial conditions and one disjoint held-out phase while preserving periodicity and the divergence-free initial velocity, and verification and weak scaling use the canonical zero-phase case.

Following a correctness-before-performance order, we first verify the assembled solver--ML gradient using Taylor remainder and centered finite-difference tests. We then evaluate multi-horizon training with fixed held-out validation, optimizer-enabled training-update weak scaling on \aurora{}, 200-step autoregressive accuracy relative to the uncorrected coarse-grid $P=2$ solver, and separate 12-rank inference wall-clock speedup relative to the corresponding $P=7$ configurations. Complete numerical and evaluation protocols are given in the Appendix.

\begin{figure}[t]
    \centering
    \includegraphics[
        width=0.9\linewidth,
        height=0.72\textheight,
        keepaspectratio
    ]{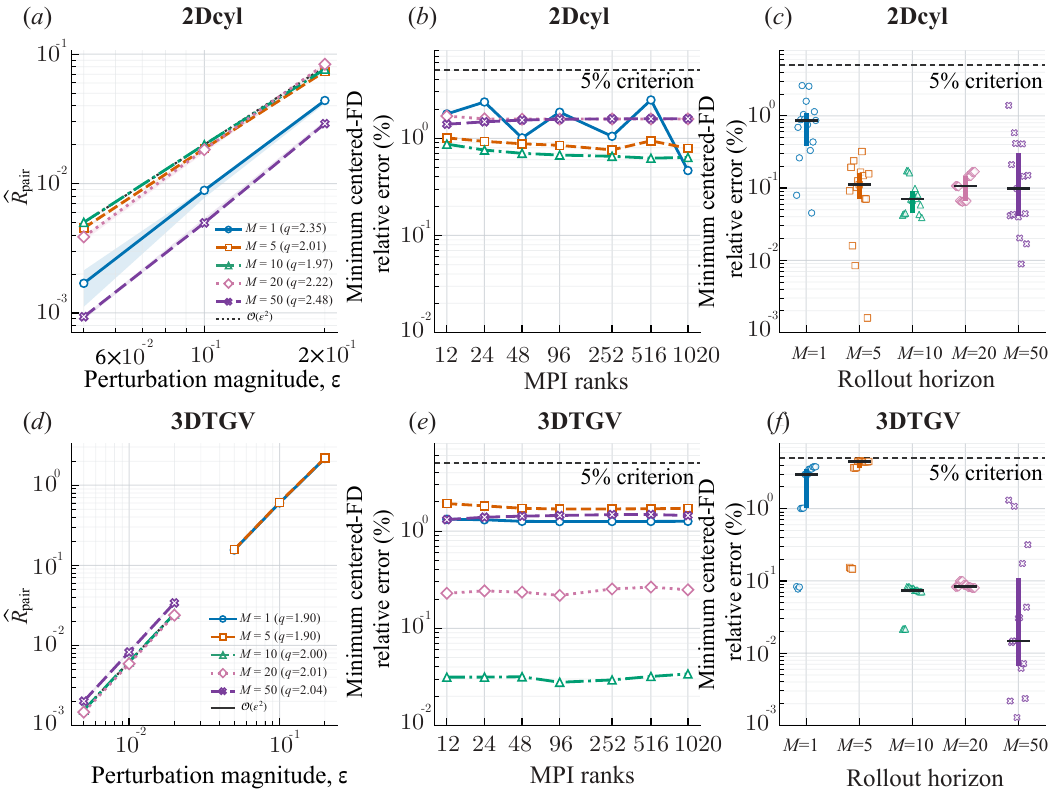}
    \caption{End-to-end exact-adjoint verification at rollout horizons $M\in\{1,5,10,20,50\}$. The top and bottom rows show 2Dcyl and 3DTGV, respectively: (\textit{a},\textit{d}) paired Taylor consistency, (\textit{b},\textit{e}) the minimum centered-FD directional error across seven weak-scaling allocations, and (\textit{c},\textit{f}) 15-direction robustness at 12 MPI ranks. In (\textit{c},\textit{f}), markers are individual directions, colored vertical bars span the interquartile range, and black horizontal bars denote the median.}
    \label{fig:e2e-gradient-verification}
\end{figure}

\subsection{End-to-End Gradient Verification}
\label{sec:results-taylor}

For each problem--horizon--allocation verification test, let $p$ be a fixed unit direction in the complete neural-parameter space. The direction is constructed as described in Appendix~\ref{app:gradient-checks} and held fixed over the perturbation sweep. For the assembled adjoint gradient $g_{\mathrm{adj}}$, we define the directional derivative $d_{\mathrm{adj}}=g_{\mathrm{adj}}^{\mathsf{T}}p$ and the paired Taylor remainder
\begin{equation}
    R_{\mathrm{pair}}(\epsilon)
    =
    \max_{\pm}
    \left|
      \mathcal{L}_M(\theta\pm\epsilon p)
      -
      \mathcal{L}_M(\theta)
      \mp
      \epsilon d_{\mathrm{adj}}
    \right|.
    \label{eq:taylor-paired}
\end{equation}
We report the normalized remainder
\begin{equation}
    \widehat{R}_{\mathrm{pair}}(\epsilon)
    =
    \frac{R_{\mathrm{pair}}(\epsilon)}
         {|d_{\mathrm{adj}}|}.
    \label{eq:normalized-taylor-paired}
\end{equation}
The normalization removes differences in directional-gradient scale without changing the convergence order. If $d_{\mathrm{adj}}$ is correct, subtracting the linear term cancels the $\mathcal{O}(\epsilon)$ objective variation and leaves $\widehat{R}_{\mathrm{pair}}=\mathcal{O}(\epsilon^2)$.
An incorrect directional derivative generally leaves an $\mathcal{O}(\epsilon)$ remainder. The Taylor test therefore checks whether the implemented adjoint correctly captures the first-order variation of the full rollout objective~\cite{farrell2013automated,griewank2008evaluating}. We adopt a prespecified 5\% engineering tolerance, consistent with the scale of adjoint--finite-difference discrepancies reported in prior CFD adjoint verification studies~\cite{anderson1999aerodynamic,wang2010adjoint}.

We additionally compare $d_{\mathrm{adj}}$ with the forward-only centered finite-difference estimate
\begin{equation}
    d_{\mathrm{FD}}(\epsilon)
    =
    \frac{
      \mathcal{L}_M(\theta+\epsilon p)
      -
      \mathcal{L}_M(\theta-\epsilon p)
    }{2\epsilon},
    \label{eq:centered-directional-derivative}
\end{equation}
using
\begin{equation}
    e_{\mathrm{FD}}(\epsilon)
    =
    \frac{
      \left|
        d_{\mathrm{FD}}(\epsilon)-d_{\mathrm{adj}}
      \right|
    }{
      \max\!\left(
        |d_{\mathrm{FD}}(\epsilon)|,
        |d_{\mathrm{adj}}|,
        \delta
      \right)
    },
    \label{eq:centered-directional-error}
\end{equation}
where $\delta$ is a small numerical floor. The Taylor test checks first-order cancellation and convergence order, whereas the centered finite difference checks the numerical magnitude and sign of the directional derivative. Every perturbed objective evaluation reruns the full authoritative \nekrs{} rollout at the specified horizon, and no fitted gradient scale is applied.

Fig.~\ref{fig:e2e-gradient-verification}(\textit{a}) and~(\textit{d}) show near-quadratic Taylor convergence. The horizontal separation in panel~(\textit{d}) reflects the prescribed 3DTGV perturbation ranges $\epsilon\in\{0.2,0.1,0.05\}$ for $M=1,5$ and $\epsilon\in\{0.02,0.01,0.005\}$ for $M=10,20,50$. The smaller range keeps the longer rollouts within the locally linear Taylor regime, while the larger short-horizon range maintains an objective signal above numerical noise. The fitted orders are $q=1.97$--$2.48$ for 2Dcyl and $q=1.90$--$2.04$ for 3DTGV. This behavior persists through $M=50$, indicating correct first-order cancellation across the full multi-timestep reverse recurrence. Fig.~\ref{fig:e2e-gradient-verification}(\textit{b}) and~(\textit{e}) show the minimum centered finite-difference error over the perturbation magnitudes considered, and that all 70
problem--horizon--allocation ($2\times 5\times 7 =70$) combinations satisfy the prespecified $5\%$ centered finite-difference criterion. Fig.~\ref{fig:e2e-gradient-verification}(\textit{c}) and~(\textit{f}) extend the representative 12-rank test to 15 directions at every rollout horizon. All 150 problem--horizon--direction ($2\times 5\times 15=150$) records satisfy the prespecified $5\%$ criterion, all accepted centered-FD signs match the adjoint signs, and the largest minimum-over-$\epsilon$ relative error is $4.56\%$. These checks broaden both the sampled state windows and parameter-space directions, but do not constitute a proof over the full parameter space.

Additionally, Appendix~\ref{app:gradient-checks} reports a controlled $M=5$ 2Dcyl comparison with the two surrogate-gradient paths used in our earlier formulation~\cite{jung2026hybrid}. The exact adjoint passes all 15 held-out directional checks and recovers near-quadratic Taylor behavior, whereas neither calibrated surrogate passes the $5\%$ gate.

Overall, these tests provide end-to-end numerical verification that the implemented adjoint computes the directional derivative of the fully discrete rollout objective for the tested flows, horizons, parameter directions, and MPI
allocations. The verification covers the assembled hybrid solver-ML gradient,
rather than only individual transpose kernels.

\subsection{Multi-timestep Training Behavior}
\label{sec:results-training}
\begin{figure}[!t]
    \centering
    \includegraphics[
        width=0.8\linewidth,
        height=0.72\textheight,
        keepaspectratio
    ]{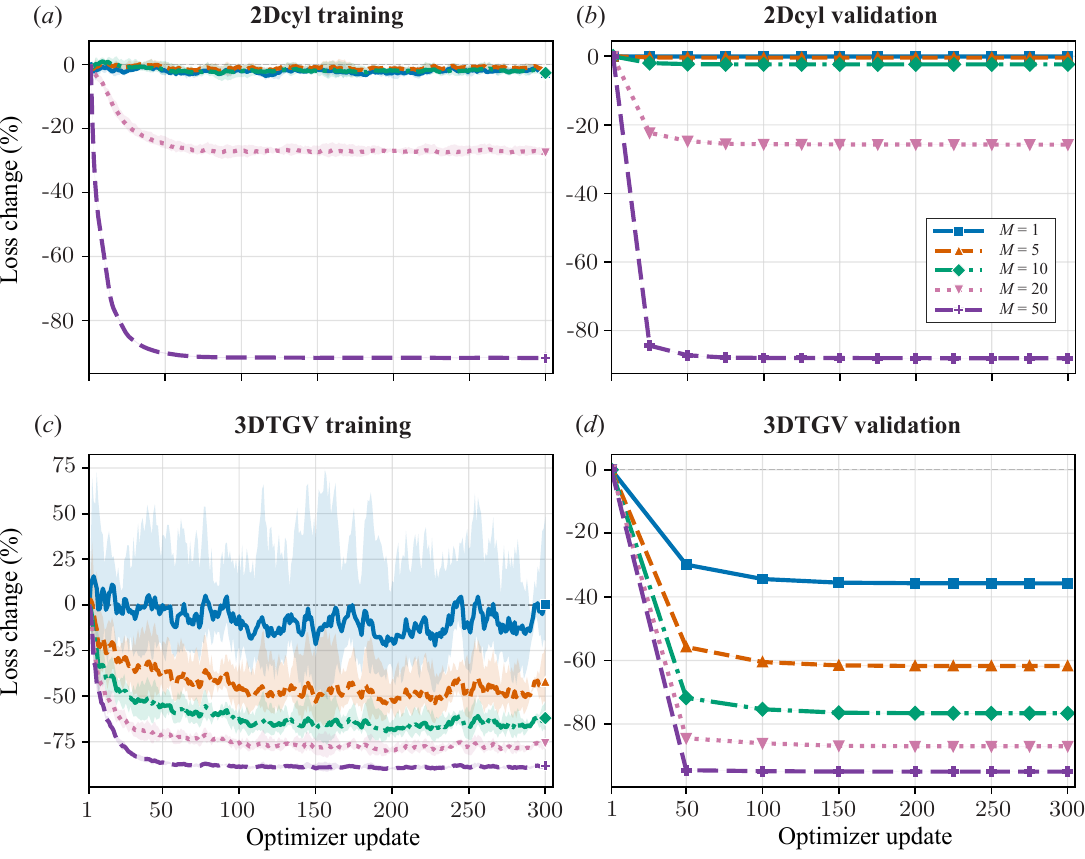}
    \caption{Multi-timestep training for (\textit{a},\textit{b}) 2Dcyl and
(\textit{c},\textit{d}) phase-shifted 3DTGV. Curves show three-seed geometric means, and shaded regions span the seedwise minimum--maximum range. Training and fixed held-out validation objectives are normalized within each run to the first optimizer update.}
    \label{fig:training-horizon}
\end{figure}
Fig.~\ref{fig:training-horizon} reports three-seed geometric means and minimum--maximum seed bands for the training and fixed
held-out validation objectives. Each curve is normalized to its corresponding value at update~1 because the rollout horizon $M$
defines a different trajectory-level objective and therefore a different absolute loss scale. The curves consequently compare relative optimization progress within each run, meaning that their vertical positions should not be interpreted as absolute loss comparisons across horizons.

The horizon dependence differs between the two flows. For 2Dcyl, the $M=50$ objective decreases rapidly during approximately the first 50 updates and then remains near a $90\%$ reduction for both training and held-out validation. The $M=20$ case produces a more moderate reduction of approximately $25$--$30\%$, whereas the $M\leq10$ objectives change only modestly. For 3DTGV, the horizon dependence is more gradual and nearly ordered by $M$. The $M=1$ training curve exhibits the largest fluctuations and seed-to-seed range, while longer horizons produce progressively larger and more consistent reductions. Validation on the held-out phase preserves this ordering, with most of the improvement occurring within the first 50--100 updates and no systematic late-iteration rebound.

All reported runs completed 300 optimizer updates with a batch size of $B=4$, and at every reported validation checkpoint after update 1, the fixed validation objective remained below its update-1 value. These results demonstrate that the verified exact gradients support optimization through $M=50$ and transfer to fixed validation windows. The inferences deployed from these models are compared in Section~\ref{sec:results-inference}. Detailed sampling, optimizer, validation, and completion controls are provided in Appendix~\ref{app:training-comparison}.
\subsection{Weak Scaling of Optimizer-Enabled Training Updates}
\label{sec:results-training-scaling}
We next characterize the weak scaling of the optimizer-enabled training path. One optimizer update comprises $B=4$ sequential
domain-parallel rollouts, averaging of the four MPI-assembled parameter gradients, unit-norm clipping, and one synchronized Adam
update while retaining the updated model and optimizer state. We test all rollout horizons $M\in\{1,5,10,20,50\}$ at all seven weak-scaling allocations. For each configuration, one complete $B=4$ update is discarded as a warm-up, and the next five complete updates are retained. To isolate the coupled training path developed in this work, we exclude the solver-reported \nekrs{} launch and setup intervals and report the remaining wall time per rollout. Weak-scaling efficiency is normalized by the corresponding 12-rank median. The complete timing and accounting protocol is provided in Appendix~\ref{app:weak-scaling}.

\begin{figure}[!t]
  \centering
  \includegraphics[
    width=0.80\linewidth,
    height=0.72\textheight,
    keepaspectratio
  ]{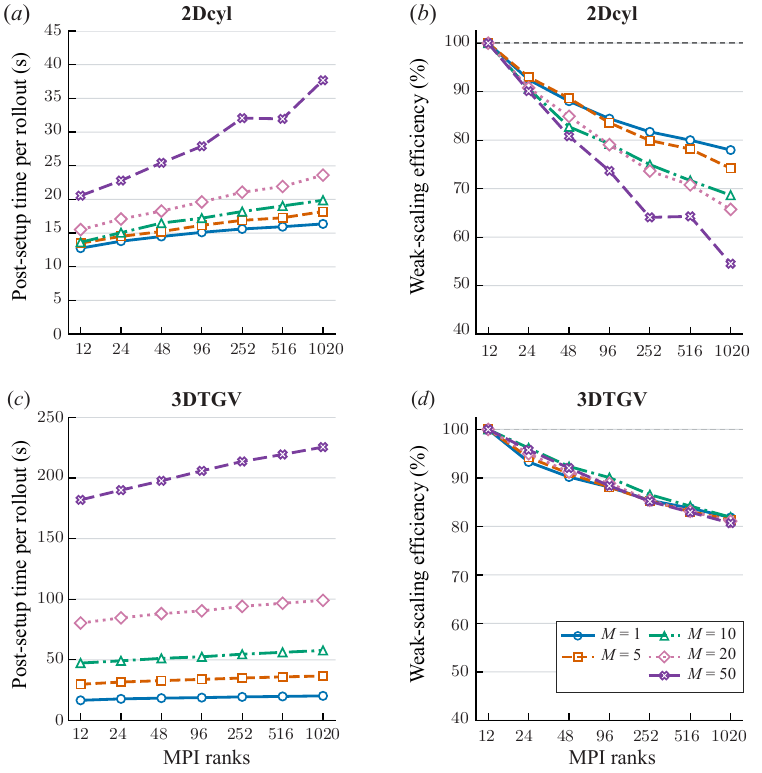}
  \caption{Optimizer-enabled training-update weak scaling on \aurora{} for (\textit{a},\textit{b}) 2Dcyl and (\textit{c},\textit{d}) 3DTGV at $M\in\{1,5,10,20,50\}$. Each marker is the median over five retained post-warm-up complete $B=4$ optimizer updates. Panels~(\textit{a},\textit{c}) show the post-setup time per rollout, and
  panels~(\textit{b},\textit{d}) show weak-scaling efficiency relative to the corresponding 12-rank median.}
  \label{fig:training-iteration-scaling}
\end{figure}

Fig.~\ref{fig:training-iteration-scaling} covers all 70 problem--horizon--allocation configurations.
At 1,020 MPI ranks, 3DTGV retains $80.7\%$--$81.9\%$ weak-scaling efficiency across the five rollout horizons. The five 3DTGV efficiency curves remain closely grouped, indicating that increasing the rollout horizon does not substantially alter the scaling behavior of this larger rank-local workload. The 2Dcyl efficiency is more horizon-dependent, decreasing from $78.0\%$ at $M=1$ to $54.5\%$ at $M=50$. This stronger degradation is consistent with the substantially smaller rank-local workload. In this work, the weak-scaling meshes contain 261 elements per rank for 2Dcyl, compared with 3,888 elements per rank for 3DTGV. The smaller 2Dcyl configuration provides less computation over which to amortize the distributed solver, adjoint, communication, and synchronization costs. We also confirmed that gradient averaging, unit-norm clipping, the Adam update, and the in-memory model and optimizer-state update require $0.420$--$1.078$~s per rollout and account for $0.368\%$--$6.151\%$ of the reported time ($B=4$) across all 70 configurations. The largest fractions occur for the $M=1$ rollouts because the nearly horizon-independent update cost is compared with only a one-step forward--reverse trajectory.

For the longest rollout at $M=50$ and 1,020 ranks, the median post-setup time per rollout is $37.695$~s for 2Dcyl and
$225.457$~s for 3DTGV, corresponding to weak-scaling efficiencies of $54.5\%$ and $80.7\%$, respectively. All 350 retained
post-warm-up complete $B=4$ observations produced finite, nonzero applied updates. As the weak-scaling mesh changes with
allocation, these experiments assess the systems execution of optimizer-enabled training updates rather than convergence of a
common learning problem across allocations. The corresponding optimizer-free exact-gradient scaling baseline is reported in Appendix~\ref{app:weak-scaling}.
\subsection{Autoregressive Inference}
\label{sec:results-inference}

We next test behavior beyond the training horizons in Section~\ref{sec:results-training}. Starting from a projected high-fidelity state, each frozen hybrid model advances for 200 consecutive timesteps and feeds its predicted state into the next solver and neural-source evaluation. The uncorrected \nekrs{} coarse-grid $P=2$ baseline uses the same solver configuration with $s_{\theta}\equiv0$. The projected high-fidelity trajectory is used only to evaluate the error after each completed inference step. We compute the mass-weighted relative velocity error
\begin{equation}
    \varepsilon_r
    =
    \frac{
        \left\|
            u_L^{k+r}-u_{\mathrm{ref}}^{k+r}
        \right\|_{M_L}
    }{
        \left\|
            u_{\mathrm{ref}}^{k+r}
        \right\|_{M_L}
    },
    \qquad
    \|v\|_{M_L}^{2}=v^{\mathsf{T}}M_Lv,
    \label{eq:inference-relative-error}
\end{equation}
where $M_L$ is the coarse-grid spectral-element mass matrix. Here, $M$ denotes the training rollout horizon and $r$ the inference step. For each training horizon $M$ and inference step $r$, we compute $\varepsilon_r$ for three independently trained optimizer seeds. We report the median of the three seedwise errors at each step, and the shaded band spans their minimum and maximum values. We present the 2Dcyl results first, followed by the 3DTGV results.

\begin{figure}[!t]
    \centering
    \includegraphics[
        width=0.9\linewidth,
        height=0.72\textheight,
        keepaspectratio
    ]{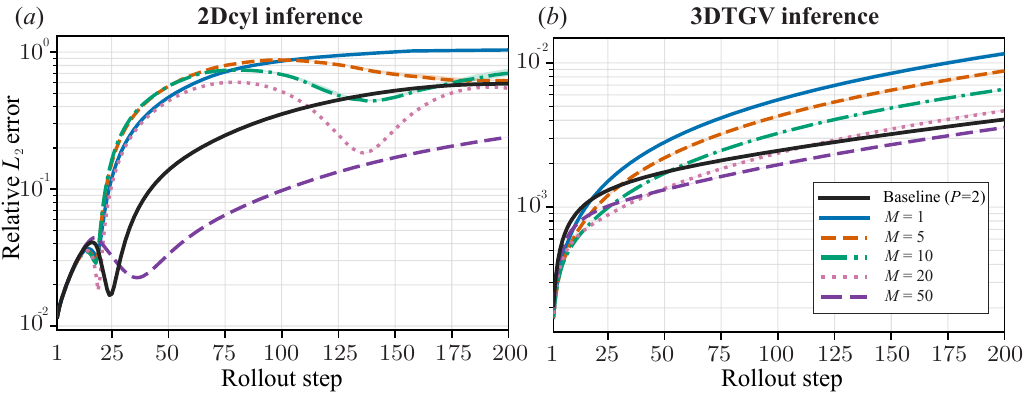}
    \caption{Relative $L_2$ velocity error over 200 fully
    autoregressive inference steps for (\textit{a}) the temporally held-out
    2Dcyl wake and (\textit{b}) a held-out 3DTGV trajectory.}
    \label{fig:inference-error}
\end{figure}

Fig.~\ref{fig:inference-error} shows that for 2Dcyl the error oscillations are consistent with accumulated phase error in the periodic vortex street. The entire $M=50$ seed band becomes more accurate than the $P=2$ baseline at step~29 and remains so through step~200. At inference steps 50, 100, and 200, its median errors are 0.0334, 0.0976, and 0.240, compared with 0.137, 0.353, and 0.588 for the baseline---reductions of 75.7\%, 72.4\%, and 59.2\%, respectively. This advantage is not monotone in $M$. At step~200, the $M=20$ median is only modestly below the baseline (0.544 versus 0.588), while the $M=1$, 5, and 10 medians are larger. Thus, exposure to a long training rollout is important for this wake, but horizon length alone does not guarantee improvement.

\begin{figure}[!t]
    \centering
    \includegraphics[
        width=0.9\linewidth,
        height=0.72\textheight,
        keepaspectratio
    ]{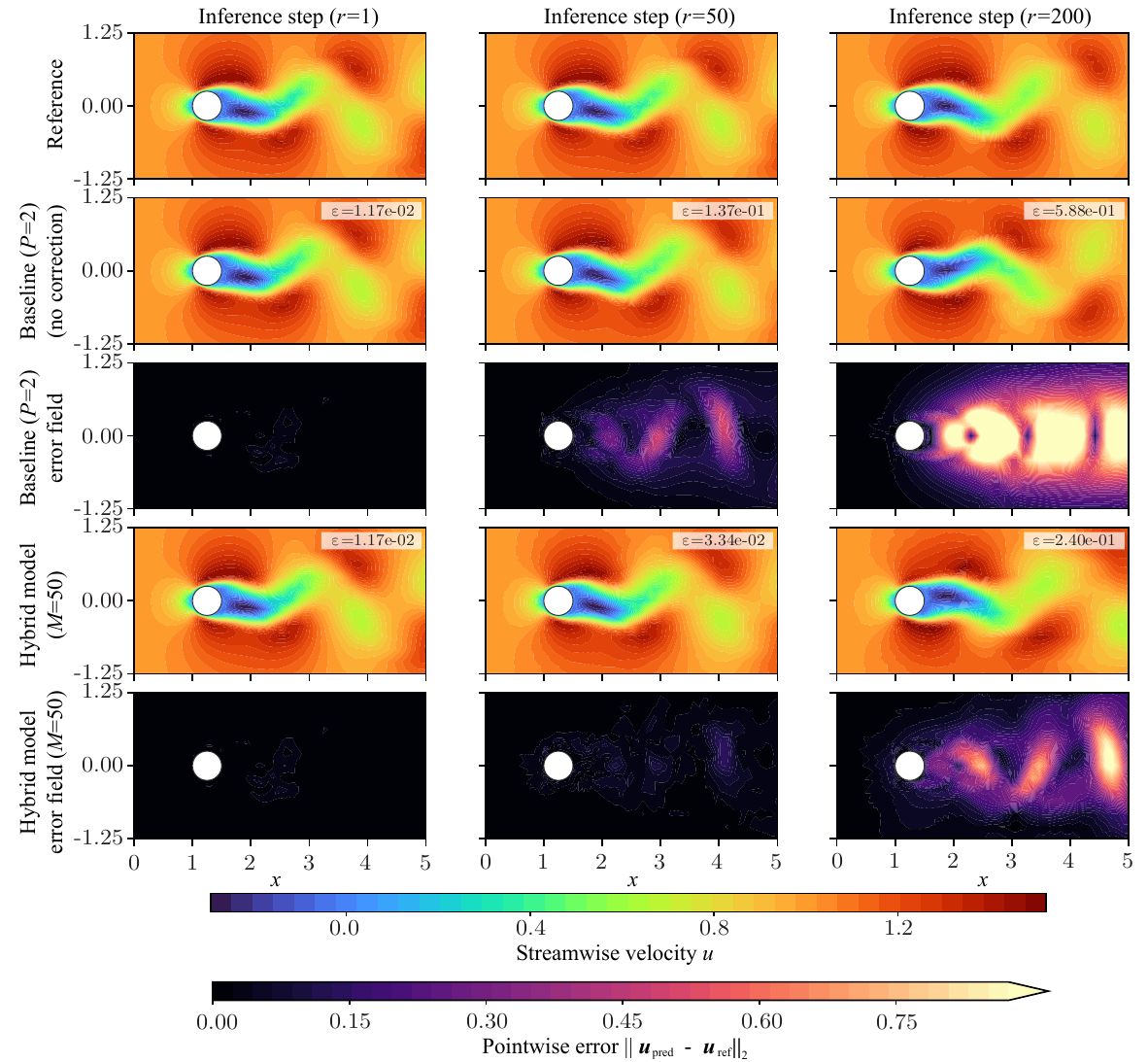}
    \caption{Spatial fields of the 2Dcyl inference and the corresponding errors at inference steps $r=1$, 50, and 200. Top to bottom: the projected high-order reference streamwise velocity $u$; the $P=2$ baseline and its pointwise error magnitude; and a representative $M=50$ prediction and its pointwise error magnitude.}
    \label{fig:inference-fields}

\end{figure}

Fig.~\ref{fig:inference-fields} clarifies how the scalar error in Fig.~\ref{fig:inference-error}(\textit{a}) develops spatially. At step~1, the
baseline and hybrid fields are both close to the projected reference, with $\varepsilon_1=1.17\times10^{-2}$. By step~50, the baseline error has spread
along the alternating near-wake structures and reaches 0.137, consistent with a growing phase and amplitude mismatch. The $M=50$ prediction remains more
closely aligned with the reference vortex street and limits the error to 0.0334. At step~200, both trajectories have accumulated error, but the
baseline exhibits a broad wake mismatch with $\varepsilon_{200}=0.588$, whereas the hybrid model retains more of the reference wake organization and reduces
the error to 0.240. The field comparison therefore indicates that long-horizon training delays the accumulation of wake phase and
amplitude errors. Fig.~\ref{fig:inference-fields} is a representative spatial realization.

In Fig.~\ref{fig:inference-error}(\textit{b}), the 3DTGV errors are smaller in magnitude and grow more smoothly. All three $M=50$ runs remain below the $P=2$ baseline at every inference step. Their
step-200 median is $3.56\times10^{-3}$, with a seed range of $[3.54,3.57]\times10^{-3}$, compared with $4.05\times10^{-3}$ for the baseline, so the median reduction is 12.1\%. Moreover, the step-200 median error decreases monotonically from $1.16\times10^{-2}$ for $M=1$ to $3.56\times10^{-3}$ for $M=50$. The present 3DTGV evaluation focuses on seed-aggregated trajectory error, so long-time physical statistics are outside the scope of this study.

Across both flows, the $M=50$ model best limits autoregressive drift, with narrow three-seed bands, although the magnitude and monotonicity of the accuracy benefit remain flow-dependent. Separate 12-rank timing using the frozen seed-1
$M=50$ model for each benchmark yields wall-clock speedups of $5.379\times$ for 2Dcyl and $2.492\times$ for 3DTGV relative to the corresponding $P=7$ configurations for equal simulated-time intervals (Appendix~\ref{app:inference-timing}). Since the configurations differ in polynomial order, BDF/EXT scheme, and, for 2Dcyl, timestep sequence, these values compare complete executed configurations rather than polynomial order alone.

\section{Conclusion}
\label{sec:conclusion}

We present Diff-NekRS, a scalable differentiable framework for multi-timestep neural-source training in the GPU-accelerated
\nekrs{} solver. The assembled solver--model gradient was verified through $M=50$ for 2Dcyl and 3DTGV across 12--1,020 MPI ranks.
The Taylor remainders were approximately quadratic, and all 70 problem--horizon--allocation checks and all
150 problem--horizon--direction checks met the prescribed $5\%$ centered finite-difference criterion. Across the tested rollout horizons, exact-adjoint training reduced every fixed held-out validation objective relative to update~1. At 1,020 MPI ranks, training with the optimizer retained a weak-scaling efficiency of 54.5\%--78.0\% for 2Dcyl and 80.7\%--81.9\% for 3DTGV. These results establish the numerical correctness and operational scalability of the Diff-NekRS training workflow for the supported fully discrete time-advancement configuration.

In 200-step autoregressive inference, the $M=50$ model reduced the three-seed median terminal relative $L_2$ velocity error by 59.2\%
for 2Dcyl and 12.1\% for 3DTGV relative to uncorrected $P=2$, and retained wall-clock speedups of $5.38\times$ and $2.49\times$, respectively, relative to the corresponding $P=7$ configurations for equal simulated-time intervals. Thus, long-horizon training improves coarse-grid trajectory accuracy while retaining a configuration-level wall-clock advantage over the high-order reference, although the trade-off remains flow-dependent. Future work will improve model efficiency and assess longer-time physical statistics, broader initial conditions, and additional NekRS time-advancement configurations.

\section*{Acknowledgment}
This work was supported by the U.S. Department of Energy Office of Science, Advanced Scientific Computing Research (ASCR), Applied Mathematics program through the Competitive Portfolios Project; the Scientific Discovery through Advanced Computing (SciDAC) FASTMath Institute; the Argonne Leadership Computing Facility, a U.S. DOE Office of Science user facility; and the Argonne Laboratory Directed Research and Development (LDRD) program, under Contract No. DE-AC02-06CH11357.

\section*{Author ORCIDs}
\orcidlink{https://orcid.org} Junoh Jung \href{https://orcid.org/0000-0003-0962-3127}{https://orcid.org/0000-0003-0962-3127};\\
\orcidlink{https://orcid.org} Riccardo Balin \href{https://orcid.org/0000-0002-1906-902X}{https://orcid.org/0000-0002-1906-902X};\\
\orcidlink{https://orcid.org} Bethany Lusch \href{https://orcid.org/0000-0002-9521-9990}{https://orcid.org/0000-0002-9521-9990};\\
\orcidlink{https://orcid.org} Emil Constantinescu \href{https://orcid.org/0000-0002-7003-6899}{https://orcid.org/0000-0002-7003-6899}.

\clearpage

\begin{appendices}
\section{Verification, Evaluation, and Reproducibility Details}
\label{app:evaluation-protocol}

\noindent
This appendix documents the benchmark configurations, exact-adjoint scope, gradient-verification procedures, training and inference controls, scaling methodology, and distributed communication and memory measurements supporting the main results. Table~\ref{tab:app-cases} summarizes the common benchmark, discretization, solver, and neural-model settings.

\begin{table}[htbp]
\centering
\caption{Benchmark, discretization, solver, and neural model settings.}
\label{tab:app-cases}
\footnotesize
\setlength{\tabcolsep}{4pt}
\renewcommand{\arraystretch}{1.12}
\begin{tabularx}{\linewidth}{@{}
  >{\raggedright\arraybackslash}p{0.14\linewidth}
  >{\raggedright\arraybackslash}X
  >{\raggedright\arraybackslash}X@{}}
\toprule
Item & 2Dcyl & 3DTGV \\
\midrule
Flow/domain
& $Re_D=100$; $[0,20D]\times[-5D,5D]\times[0,0.1D]$; cylinder at $(2.5D,0)$
& $Re=1600$; triply periodic $[-\pi,\pi]^3$ \\
Boundary/initial data
& Uniform inflow; no-slip cylinder; open outlet; free-slip lateral/spanwise faces
& $\mathbf{u}_0=(\sin x\cos y\cos z,$\newline
$\phantom{\mathbf{u}_0=(}{-}\cos x\sin y\cos z,0)$ \\
Element mesh
& $3{,}132$ hexes: $1{,}044$ cross-sectional elements $\times$ three layers
& $36^3=46{,}656$ regular hexes \\
Element-local GLL points
& $P=7$: $1{,}603{,}584$; $P=2$: $84{,}564$
& $P=7$: $23{,}887{,}872$; $P=2$: $1{,}259{,}712$ \\
Reference / differentiated step
& BDF2/EXT2 / BDF1/EXT1; target CFL~2; $\Delta t\le2\times10^{-2}$
& BDF3/EXT3 / BDF1/EXT1; $\Delta t=1.5\times10^{-3}$ for verification/scaling \\
Linear-solver tolerance
& \multicolumn{2}{>{\raggedright\arraybackslash}p{0.78\linewidth}@{}}{$10^{-10}$ relative residual tolerance for velocity and pressure in verification/scaling} \\
Neural source model
& 5-element in-plane stencil; $270\!\rightarrow\!256\!\rightarrow\!256\!\rightarrow\!54$; $149{,}046$ parameters
& 7-element periodic stencil; $567\!\rightarrow\!256\!\rightarrow\!256\!\rightarrow\!81$; $232{,}017$ parameters \\
Executed transform
& \multicolumn{2}{>{\raggedright\arraybackslash}p{0.78\linewidth}@{}}{Componentwise input normalization, clipping to $[-10,10]$, two GELU hidden layers, bounded $\tanh$ output, and distributed componentwise zero-mean projection} \\
Machine precision
& \multicolumn{2}{>{\raggedright\arraybackslash}p{0.78\linewidth}@{}}{NekRS state/adjoint: FP64; LibTorch model and projected targets: FP32}\\
\bottomrule
\end{tabularx}

\smallskip
\par\noindent\footnotesize\raggedright
Within each benchmark, $P=7$ reference and $P=2$ differentiated runs use the
same element topology; GLL totals are element-local and duplicate interface
points. Benchmark choices follow
Refs.~\cite{posdziech2007cylinder,williamson1996vortex,
brachet1983tgv,barwey2025meshsr,jung2026hybrid}.
\end{table}

\subsection{Exact-Adjoint Scope and Gradient Verification}
\label{app:gradient-checks}

The exactness claim is restricted to the executed BDF1/EXT1 map with cubature advection and no subcycling. The reverse path includes the neural source and zero-mean projection, boundary constraints, pressure projection, nonlinear advection, and the configured elliptic solves. Here, ``exact'' denotes algebraic transpose-Jacobian actions of this map up to Krylov-solver and floating-point tolerances. Adaptive time-stepping logic, higher-order BDF/EXT schemes, subcycling, and other \nekrs{} configurations are outside the scope of the present claim and will be considered in future work.

For each problem--horizon--allocation, the unperturbed objective is repeated
and accepted only when
\begin{equation}
|\mathcal{L}_M-\mathcal{L}_M^{\mathrm{repeat}}|
\leq\max\!\left(10^{-10},10^{-7}
\max(|\mathcal{L}_M|,|\mathcal{L}_M^{\mathrm{repeat}}|)\right).
\label{eq:app-repeatability}
\end{equation}
Here, $\mathcal{L}_M^{\mathrm{repeat}}$ denotes an independently repeated evaluation of the unperturbed $M$-step rollout objective using the same parameter vector, restart state, rollout horizon, and MPI allocation as $\mathcal{L}_M$. A deterministic unit direction is generated from a seeded random vector and the normalized adjoint, with $g_{\mathrm{adj}}^Tp/\|g_{\mathrm{adj}}\|_2=1/\sqrt{2}$, to avoid a numerically unresolved near-orthogonal direction. Every perturbed objective is a fresh authoritative rollout.

The prescribed perturbations are $\epsilon\in\{0.2,0.1,0.05\}$ for all 2Dcyl tests and for 3DTGV at $M=1,5$, and $\epsilon\in\{0.02,0.01,0.005\}$ for 3DTGV at $M=10,20,50$. The Taylor plots show the median and interquartile range over seven allocations, and the fitted order is the least-squares slope through the three median remainders. A configuration passes when $\min_\epsilon e_{\mathrm{FD}}(\epsilon)\le5\%$. This threshold is a numerical gate rather than a confidence interval. As noted in the main text, we use a prespecified 5\% engineering tolerance, consistent with the scale of adjoint--finite-difference discrepancies reported in prior CFD adjoint verification studies~\cite{anderson1999aerodynamic,wang2010adjoint}.

For each flow and \(M\in\{1,5,10,20\}\), the 15-direction robustness study evaluates three independently seeded gradient-informed directions at each of five fixed restart windows using a common problem- and horizon-specific perturbation set. For \(M=50\), it evaluates one gradient-informed direction and 14 independent unit-normalized standard-Gaussian directions (seeds 202608081--202608094) at a single representative restart window. All tests use prespecified local-regime perturbation grids and a deterministic noise-floor re-evaluation rule. Thus, the \(M\leq20\) tests vary both restart window and direction, whereas the \(M=50\) tests vary direction at a fixed window.

We recomputed the complete $M=5$ 2Dcyl parameter gradient from the same checkpoint using the exact adjoint and two frozen-state surrogate paths~\cite{jung2026hybrid}. The global linearized Navier-Stokes operator is a frozen-state, pressure-coupled linearized Navier-Stokes adjoint over the full distributed domain. Signed scalar calibration used 45 restart windows (135 directions) and was frozen before testing on five held-out windows (15 directions). All paths used the same six perturbation magnitudes.

\begin{table}[t]
  \caption{Controlled exact-versus-surrogate gradient comparison. Error is relative to centered finite differences.}
  \label{tab:exact-vs-surrogate}
  \centering
  \small
  \setlength{\tabcolsep}{4pt}
  \begin{tabular}{@{}lrrrr@{}}
    \toprule
    Gradient path & Median error & Within 5\% & Taylor $q$ & Time (s) \\
    \midrule
    Exact adjoint     & $0.103\%$ & 15/15 & 2.072 & 38.26 \\
    Block-Oseen       & $22.9\%$  & 0/15  & 1.107 & 38.99 \\
    Global linearized NS        & $52.0\%$  & 0/15  & 1.054 & 107.39 \\
    \bottomrule
  \end{tabular}
\end{table}

Times are medians of five retained one-node, 12-rank launcher-to-completion measurements after one discarded warm-up and include \nekrs{} launch/setup. In Table~\ref{tab:exact-vs-surrogate}, the exact path is within $1.9\%$ of the block-Oseen cost and $2.81\times$ faster than the global linearized NS path. This is a controlled gradient-fidelity ablation with an evaluation-cost comparison, not an attribution of the historical end-to-end training difference solely to the gradient method.
\subsection{Training and Validation Protocol}
\label{app:training-comparison}

Table~\ref{tab:app-training-controls} summarizes the training and validation controls. All displayed training runs use one Aurora node and 12 MPI ranks. Four sequential, domain-parallel rollout gradients form one effective batch. Their MPI-reduced parameter gradients are averaged, clipped to unit norm, and used for one Adam update. Uniform rollout weights $w_j=1/M$ remove the trivial factor associated with summing $M$ loss terms. Adam uses $\varepsilon_{\mathrm{Adam}}=10^{-10}$ and no weight decay.

\begin{table}[t]
\caption{Training and validation controls used for
Fig.~\ref{fig:training-horizon}.}
\label{tab:app-training-controls}
\centering
\small
\setlength{\tabcolsep}{4pt}
\renewcommand{\arraystretch}{1.12}
\begin{tabularx}{\linewidth}{@{}
  >{\raggedright\arraybackslash}p{0.19\linewidth}
  >{\raggedright\arraybackslash}X
  >{\raggedright\arraybackslash}X@{}}
\toprule
Control & 2Dcyl & 3DTGV \\
\midrule
Training population
& 500 starts from one developed trajectory after reserving five temporal validation intervals
& 1,000 starts distributed across 10 phase-shifted initial conditions \\
Sampling
& Four temporal strata; deterministic stream shared across horizons within each seed
& Four temporal strata and four distinct phases per update \\
Seeds / budget
& Three optimizer seeds; 300 completed updates per displayed horizon
& Three optimizer seeds; 300 completed updates per displayed horizon \\
Learning rate
& Cosine $10^{-3}\!\rightarrow\!10^{-6}$
& Cosine $10^{-3}\!\rightarrow\!10^{-6}$ \\
Validation
& Five same-trajectory temporal holdouts; update~1 and every 25 updates
& Held-out phase at starts $0,250,500,750,999$; prescribed checkpoints through update~300 \\
\bottomrule
\end{tabularx}
\end{table}

The 2Dcyl loss includes all GLL points belonging to elements whose centroids lie within $x/D\in(2,8)$ and $y/D\in(-2,2)$ and is normalized by active volume. 3DTGV uses the whole-domain loss and $\Delta t=3\times10^{-3}$ for the phase-family training campaign. Validation freezes the model and optimizer and performs no reverse sweep. The 2Dcyl holdouts are disjoint in time but belong to the same trajectory. The 3DTGV holds out an independently sampled phase. Fig.~\ref{fig:training-horizon} therefore reports
optimization progress and loss reduction on fixed held-out validation sets, rather than an a posteriori physical-accuracy comparison with the coarse-grid $P=2$ baseline.

\subsection{Inference Timing and Wall-Clock Speedup}
\label{app:inference-timing}

To complement the trajectory-accuracy evaluation in Section~\ref{sec:results-inference}, we measured forward inference on one \aurora{} node using 12 MPI ranks. Initialization, model loading and JIT compilation, diagnostic I/O, and field output were excluded, whereas all solver operations and neural-source evaluations were included. Hybrid timings use the frozen $M=50$ checkpoint obtained from the training campaign described in Section~\ref{sec:results-training}.

For each retained run, we define
\begin{equation}
  \mathcal{T}_{\mathrm{phys}}
  =
  \frac{\sum_n \Delta t_n}
       {\sum_n t_{\mathrm{wall},n}},
  \qquad
  S_{\mathrm{wall},P7}
  =
  \frac{\mathcal{T}_{\mathrm{phys},\mathrm{method}}}
       {\mathcal{T}_{\mathrm{phys},P=7}},
  \label{eq:app-inference-wall-speedup}
\end{equation}
where $\mathcal{T}_{\mathrm{phys}}$ is the simulated physical time advanced per wall-clock second and $S_{\mathrm{wall},P7}$ is the wall-clock speedup for advancing an equal simulated-time interval. The plain $P=2$ and hybrid $P=2 + \text{ML}$ runs use BDF1/EXT1, whereas the $P=7$ references use BDF2/EXT2 for 2Dcyl and BDF3/EXT3 for 3DTGV. The 2Dcyl speedups account for the different timestep sequences, whereas all 3DTGV runs use $\Delta t=0.003$.

\begin{table}[!t]
  \centering
  \caption{Twelve-rank forward-inference performance relative to the
  corresponding $P=7$ configurations. Values are medians over three
  retained post-warm-up runs.}
  \label{tab:app-inference-timing-diagnostics}
  \small
  \setlength{\tabcolsep}{4pt}
  \renewcommand{\arraystretch}{1.12}
  \begin{tabular*}{\linewidth}{@{\extracolsep{\fill}}llrrrr@{}}
    \toprule
    Case & Method &
    \multicolumn{1}{c}{\shortstack{Median\\$\Delta t$}} &
    \multicolumn{1}{c}{$\mathrm{CFL}_{\max}$} &
    \multicolumn{1}{c}{\shortstack{$\mathcal{T}_{\mathrm{phys}}$\\
                                     (sim.\ time/s)}} &
    \multicolumn{1}{c}{\shortstack{$S_{\mathrm{wall},P7}$\\ ($\times$)}} \\
    \midrule
    2Dcyl & Plain low-order ($P=2$)        & 0.0200 & 0.898 & 1.6175 & 6.471 \\
           & Hybrid model ($P=2+\mathrm{ML}$) & 0.0200 & 0.899 & 1.3448 & 5.379 \\
           & High-order ($P=7$)             & 0.0060 & 2.090 & 0.2500 & 1.000 \\
    \midrule
    3DTGV & Plain low-order ($P=2$)        & 0.0030 & 0.035 & 0.5005 & 5.233 \\
           & Hybrid model ($P=2+\mathrm{ML}$) & 0.0030 & 0.035 & 0.2383 & 2.492 \\
           & High-order ($P=7$)             & 0.0030 & 0.269 & 0.0957 & 1.000 \\
    \bottomrule
  \end{tabular*}
\end{table}

Table~\ref{tab:app-inference-timing-diagnostics} shows that the hybrid retains wall-clock speedups of $5.379\times$ for 2Dcyl and $2.492\times$ for 3DTGV relative to the corresponding $P=7$ configurations. The larger relative ML cost for 3DTGV is consistent with its larger rank-local workload, seven-element (3D) rather than five-element (2D) stencil, and larger neural and halo-communication workloads.

As the configurations differ in polynomial order, BDF/EXT scheme, and, for 2Dcyl, timestep sequence, the reported values compare
complete executed configurations rather than polynomial order alone. The adaptive-timestep 2Dcyl measurements are forward-only diagnostics. Together with Section~\ref{sec:results-inference}, these results show that the hybrid improves the plain $P=2$ trajectory while retaining a wall-clock advantage over the corresponding $P=7$ run.

\subsection{Weak-Scaling and Timing Protocol}
\label{app:weak-scaling}

An \aurora{} compute node contains two Intel Xeon CPU Max Series processors and six Intel Data Center GPU Max 1550 accelerators. Each physical GPU package comprises two independently addressable GPU tiles, giving 12 GPU tiles per node~\cite{allen2025aurora,alcf2026auroraoverview}. We expose the tiles as accelerator devices and launch 12 MPI ranks per node,
binding one rank to each tile. Thus, two MPI ranks are assigned to each physical GPU package, but each rank operates on a distinct GPU tile.

The weak-scaling campaign uses $N_r\in\{12,24,48,96,252,516,1020\}$ MPI ranks on $\{1,2,4,8,21,43,85\}$ nodes, respectively. The rank-to-tile mapping is held fixed across the campaign. 3DTGV uses fixed-domain $h$-refinement from $36^3$ to $153\times160\times162$ elements, maintaining 3,888 elements per rank. 2Dcyl fixes the 1,044-element cross-section and increases the spanwise extrusion from 3 to 255 layers, maintaining 261 elements per rank.

For the optimizer-enabled study in Fig.~\ref{fig:training-iteration-scaling}, each configuration discards one complete $B=4$ warm-up update and retains the next five complete updates. Each update contains four sequential domain-parallel forward--reverse rollouts, averaging of the four MPI-assembled parameter gradients, unit-norm clipping, and one synchronized Adam update, including the in-memory model and optimizer state update. To isolate the coupled training path developed in this work, the four solver-reported \nekrs{} launch and setup intervals are excluded from each retained update, and the remaining wall time is divided by $B=4$ to obtain the post-setup time per rollout. The reported value is the median over the five retained updates. At $M=50$ and $N_r=1020$, the excluded launch/setup interval accounted for approximately 52$\%$ and 23$\%$ of the complete ($B=4$) iteration time for 2Dcyl and 3DTGV, respectively. The excluded solver launch and initialization path is outside the coupled training implementation evaluated here; reducing this overhead is left to future work. Each case--horizon pair uses its corresponding 12-rank median as the weak-scaling baseline. These values are derived from complete $B=4$ updates and are not separately executed $B=1$ timings.

\begin{figure}[t]
  \centering
  \includegraphics[
    width=0.75\linewidth,
    height=0.72\textheight,
    keepaspectratio
  ]{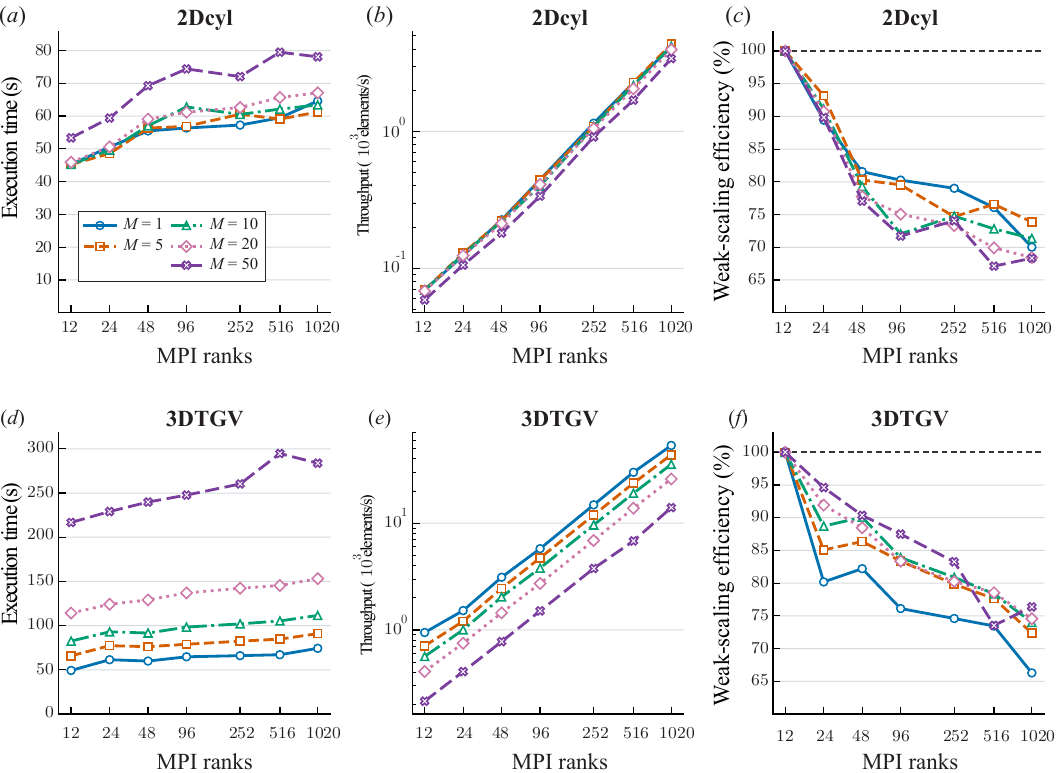}
  \caption{Optimizer-free weak scaling of one full forward--reverse
  exact-gradient evaluation on \aurora{} for
  (\textit{a}--\textit{c}) 2Dcyl and
  (\textit{d}--\textit{f}) 3DTGV. Markers denote medians over three post-warm-up trials. Throughput and weak-scaling efficiency are computed from the median times and normalized by the corresponding 12-rank result.}
  \label{fig:app-exact-gradient-weak-scaling}
\end{figure}

For comparison, we separately measure one full, unperturbed forward--reverse exact-gradient evaluation with the optimizer disabled. These measurements include the forward rollout, reverse replay, distributed halo operations, neural-model VJPs, and global parameter-gradient assembly. For each problem--horizon--allocation configuration, three repeated trials are retained after warm-up under the same rank-to-tile binding, and Fig.~\ref{fig:app-exact-gradient-weak-scaling} reports their median. Each problem-horizon pair uses its 12-rank median as its baseline.

We define the aggregate element throughput and
weak-scaling efficiency as
\begin{align}
    Q(N_r)
    &=
    \frac{N_{\mathrm{elem}}(N_r)}
         {T(N_r)},
    \\
    \eta_{\mathrm{weak}}(N_r;N_{r,0})
    &=
    \frac{T(N_{r,0})}{T(N_r)}
    \times 100\%,
    \label{eq:weak-scaling-efficiency}
\end{align}
where $T(N_r)$ denotes the selected post-warm-up wall time on $N_r$ MPI ranks, and $N_{r,0}=12$ denotes the baseline allocation. Here, $N_{\mathrm{elem}}(N_r)$ is the global low-order ($P=2$) element count. The aggregate-throughput increases are $58.1$--$62.8\times$ for 2Dcyl and $56.3$--$64.9\times$ for 3DTGV. At 1,020 ranks, the corresponding optimizer-free exact-gradient efficiencies are $68.3\%$--$73.9\%$ for 2Dcyl and $66.3\%$--$76.4\%$ for 3DTGV, with median wall times of $61.2$--$78.1$~s and $74.3$--$283.7$~s, respectively. Since it uses the full exact-gradient-evaluation timing boundary, whereas Fig.~\ref{fig:training-iteration-scaling} reports setup-excluded time per rollout derived from complete $B=4$ updates, the two timing sets are complementary rather than directly subtractable.

\subsection{Halo-Transpose Verification and Tape Storage}
\label{app:halo-details}

Let $H$ denote the forward halo gather that maps owned element data to the off-rank rows required by the neural stencil, and let $H^{\mathsf{T}}$ denote the corresponding reverse scatter-add. For deterministic test vectors $u$ and $z$, where $u$
has the same distributed layout as the rank-owned element data supplied to the forward gather and $z$ has the same layout as the gathered off-rank halo rows, we define the relative dot-product error
\begin{equation}
    e_H
    =
    \frac{
      \left|
        \langle Hu,z\rangle
        -
        \langle u,H^{\mathsf{T}}z\rangle
      \right|
    }{
      \max\!\left(
        |\langle Hu,z\rangle|,
        |\langle u,H^{\mathsf{T}}z\rangle|,
        \delta
      \right)
    },
    \label{eq:app-halo-transpose-error}
\end{equation}
where $\delta>0$ is a small numerical floor. This test verifies that the reverse communication applies the transpose of the forward halo mapping, including accumulation of received sensitivities into their owning elements. Table~\ref{tab:halo-transpose-results} reports the resulting errors.  Overall, every allocation satisfies the prescribed transpose-consistency test.

\begin{table}[t]
  \centering
  \caption{Distributed halo-transpose dot-product error across the
  weak-scaling allocations.}
  \label{tab:halo-transpose-results}
  \small
  \setlength{\tabcolsep}{5.0pt}
  \begin{tabular}{@{}rrrr@{}}
    \toprule
    Nodes & Ranks & 2Dcyl $e_H$ & 3DTGV $e_H$ \\
    \midrule
     1 &   12 & $1.64\times10^{-6}$ & $6.67\times10^{-9}$ \\
     2 &   24 & $7.95\times10^{-7}$ & $1.56\times10^{-8}$ \\
     4 &   48 & $3.95\times10^{-7}$ & $2.69\times10^{-8}$ \\
     8 &   96 & $2.25\times10^{-7}$ & $3.37\times10^{-8}$ \\
    21 &  252 & $6.45\times10^{-8}$ & $2.69\times10^{-8}$ \\
    43 &  516 & $5.34\times10^{-8}$ & $2.18\times10^{-8}$ \\
    85 & 1020 & $1.34\times10^{-8}$ & $3.24\times10^{-8}$ \\
    \bottomrule
  \end{tabular}
\end{table}
At 1,020 ranks, one halo action carries 30.23~MB and 511.02~MB of aggregate application data for 2Dcyl and 3DTGV, respectively. An $M$-step exact-gradient evaluation performs $2M$ forward gathers and $M$ transpose scatters. Each rank stores $M+1$ model-facing and solver-velocity states together with $M$ timestep values, giving an explicit tape-storage complexity of $\mathcal{O}(M N_{\mathrm{local}})$. At $M=50$, the rank-local XPU tape requires 4.11~MiB/rank for 2Dcyl and 61.27~MiB/rank for 3DTGV, well below the nominal 64-GB tile-local HBM capacity. These memory measurements are limited to one node.

\end{appendices}

\clearpage
\bibliographystyle{unsrtnat}
\bibliography{references}

@misc{alcf2026auroraoverview,
  author       = {{Argonne Leadership Computing Facility}},
  title        = {{Aurora} Machine Overview},
  year         = {2026},
  howpublished = {\url{https://docs.alcf.anl.gov/aurora/}},
  note         = {Accessed: Aug. 8, 2026}
}

@article{allen2025aurora,
  author        = {Allcock, William E. and Allen, Benjamin S. and Anchell, James and Anisimov, Victor and others},
  title         = {{Aurora}: Architecting {Argonne}'s First Exascale Supercomputer for Accelerated Scientific Discovery},
  journal       = {arXiv preprint arXiv:2509.08207},
  year          = {2025},
  eprint        = {2509.08207},
  archivePrefix = {arXiv},
  primaryClass  = {cs.DC},
  doi           = {10.48550/arXiv.2509.08207}
}

@article{anderson1999aerodynamic,
  author  = {Anderson, W. Kyle and Venkatakrishnan, V.},
  title   = {Aerodynamic Design Optimization on Unstructured
             Grids with a Continuous Adjoint Formulation},
  journal = {Computers \& Fluids},
  volume  = {28},
  number  = {4--5},
  pages   = {443--480},
  year    = {1999},
  doi     = {10.1016/S0045-7930(98)00041-3}
}

@inproceedings{chen2018neuralode,
  author        = {Chen, Ricky T. Q. and Rubanova, Yulia and Bettencourt, Jesse and
                   Duvenaud, David},
  title         = {Neural Ordinary Differential Equations},
  booktitle     = {Advances in Neural Information Processing Systems},
  volume        = {31},
  year          = {2018}
}

@article{farrell2013automated,
  author        = {Farrell, Patrick E. and Ham, David A. and Funke, Simon W. and
                   Rognes, Marie E.},
  title         = {Automated Derivation of the Adjoint of High-Level Transient Finite
                   Element Programs},
  journal       = {SIAM Journal on Scientific Computing},
  volume        = {35},
  number        = {4},
  pages         = {C369--C393},
  year          = {2013},
  doi           = {10.1137/120873558}
}

@article{giles2000adjoint,
  author        = {Giles, Michael B. and Pierce, Niles A.},
  title         = {An Introduction to the Adjoint Approach to Design},
  journal       = {Flow, Turbulence and Combustion},
  volume        = {65},
  number        = {3--4},
  pages         = {393--415},
  year          = {2000},
  doi           = {10.1023/A:1011430410075}
}

@book{griewank2008evaluating,
  author        = {Griewank, Andreas and Walther, Andrea},
  title         = {Evaluating Derivatives: Principles and Techniques of Algorithmic
                   Differentiation},
  edition       = {2},
  publisher     = {Society for Industrial and Applied Mathematics},
  address       = {Philadelphia, PA, USA},
  year          = {2008},
  doi           = {10.1137/1.9780898717761}
}

@inproceedings{hu2020difftaichi,
  author        = {Hu, Yuanming and Anderson, Luke and Li, Tzu-Mao and Sun, Qi and
                   Carr, Nathan and Ragan-Kelley, Jonathan and Durand, Fr{\'e}do},
  title         = {{DiffTaichi}: Differentiable Programming for Physical Simulation},
  booktitle     = {International Conference on Learning Representations},
  year          = {2020}
}

@article{bezgin2025jaxfluids2,
  author  = {Bezgin, Deniz A. and Buhendwa, Aaron B. and Adams, Nikolaus A.},
  title   = {{JAX-Fluids} 2.0: Towards {HPC} for Differentiable {CFD} of
             Compressible Two-Phase Flows},
  journal = {Computer Physics Communications},
  volume  = {308},
  pages   = {109433},
  year    = {2025},
  doi     = {10.1016/j.cpc.2024.109433}
}

@article{franz2026pict,
  author  = {Franz, Aleksandra and Wei, Hao and Guastoni, Luca and Thuerey, Nils},
title = {{PICT}--A Differentiable, {GPU}-Accelerated Multi-Block
         {PISO} Solver for Simulation-Coupled Learning Tasks
         in Fluid Dynamics},
  journal = {Journal of Computational Physics},
  volume  = {544},
  pages   = {114433},
  year    = {2026},
  doi     = {10.1016/j.jcp.2025.114433}
}

@article{weymouth2025waterlily,
  author  = {Weymouth, Gabriel D. and Font, Bernat},
  title   = {{WaterLily.jl}: A Differentiable and Backend-Agnostic {Julia}
             Solver for Incompressible Viscous Flow Around Dynamic Bodies},
  journal = {Computer Physics Communications},
  volume  = {315},
  pages   = {109748},
  year    = {2025},
  doi     = {10.1016/j.cpc.2025.109748}
}

@article{rackauckas2019diffeqflux,
  author        = {Rackauckas, Christopher and Innes, Mike and Ma, Yingbo and
                   Bettencourt, Jesse and White, Lyndon and Dixit, Vaibhav},
  title         = {{DiffEqFlux.jl}: A {Julia} Library for Neural Differential Equations},
  journal       = {arXiv preprint arXiv:1902.02376},
  year          = {2019},
  eprint        = {1902.02376},
  archivePrefix = {arXiv},
  doi           = {10.48550/arXiv.1902.02376}
}

@article{rackauckas2020universal,
  author        = {Rackauckas, Christopher and Ma, Yingbo and Martensen, Julius and
                   Warner, Collin and Zubov, Kirill and Supekar, Rohit and Skinner,
                   Dominic and Ramadhan, Ali and Edelman, Alan},
  title         = {Universal Differential Equations for Scientific Machine Learning},
  journal       = {arXiv preprint arXiv:2001.04385},
  year          = {2020},
  eprint        = {2001.04385},
  archivePrefix = {arXiv},
  doi           = {10.48550/arXiv.2001.04385}
}

@article{wang2010adjoint,
  author  = {Wang, D. X. and He, L.},
  title   = {Adjoint Aerodynamic Design Optimization for Blades
             in Multistage Turbomachines---{Part I}:
             Methodology and Verification},
  journal = {Journal of Turbomachinery},
  volume  = {132},
  number  = {2},
  pages   = {021011},
  year    = {2010},
  doi     = {10.1115/1.3072498}
}

@article{fan2025neuraldifferentiable,
  author        = {Fan, Xiantao and Akhare, Deepak and Wang, Jian-Xun},
  title         = {Neural Differentiable Modeling with Diffusion-Based
                   Super-Resolution for Two-Dimensional Spatiotemporal Turbulence},
  journal       = {Computer Methods in Applied Mechanics and Engineering},
  volume        = {433},
  pages         = {117478},
  year          = {2025},
  doi           = {10.1016/j.cma.2024.117478}
}

@article{fan2026coupledclosures,
  author        = {Fan, Xiantao and Liu, Yi and Wang, Meng and Wang, Jian-Xun},
  title         = {Differentiable Hybrid Neural--{CFD} Modelling of Wall-Bounded
                   Turbulence: Coupled Learning of Subgrid-Scale and Wall Closures},
  journal       = {arXiv preprint arXiv:2607.17357},
  year          = {2026},
  eprint        = {2607.17357},
  archivePrefix = {arXiv},
  primaryClass  = {physics.flu-dyn},
  doi           = {10.48550/arXiv.2607.17357}
}

@article{kochkov2021mlcfd,
  author        = {Kochkov, Dmitrii and Smith, Jamie A. and Alieva, Ayya and Wang,
                   Qing and Brenner, Michael P. and Hoyer, Stephan},
  title         = {Machine Learning--Accelerated Computational Fluid Dynamics},
  journal       = {Proceedings of the National Academy of Sciences},
  volume        = {118},
  number        = {21},
  pages         = {e2101784118},
  year          = {2021},
  doi           = {10.1073/pnas.2101784118}
}

@article{list2022learnedturbulence,
  author        = {List, Bj{\"o}rn and Chen, Li-Wei and Thuerey, Nils},
  title         = {Learned Turbulence Modelling with Differentiable Fluid Solvers:
                   Physics-Based Loss Functions and Optimisation Horizons},
  journal       = {Journal of Fluid Mechanics},
  volume        = {949},
  pages         = {A25},
  year          = {2022},
  doi           = {10.1017/jfm.2022.738}
}

@article{macart2021embedded,
  author        = {MacArt, Jonathan F. and Sirignano, Justin A. and Freund, Jonathan
                   B.},
  title         = {Embedded Training of Neural-Network Subgrid-Scale Turbulence Models},
  journal       = {Physical Review Fluids},
  volume        = {6},
  number        = {5},
  pages         = {050502},
  year          = {2021},
  doi           = {10.1103/PhysRevFluids.6.050502}
}

@article{maulik2019subgrid,
  author        = {Maulik, Romit and San, Omer and Rasheed, Adil and Vedula, Prakash},
  title         = {Subgrid Modelling for Two-Dimensional Turbulence Using Neural
                   Networks},
  journal       = {Journal of Fluid Mechanics},
  volume        = {858},
  pages         = {122--144},
  year          = {2019},
  doi           = {10.1017/jfm.2018.770}
}

@article{shankar2025differentiableturbulence,
  author        = {Shankar, Varun and Chakraborty, Dibyajyoti and Viswanathan,
                   Venkatasubramanian and Maulik, Romit},
  title         = {Differentiable Turbulence: Closure as a Partial Differential
                   Equation Constrained Optimization},
  journal       = {Physical Review Fluids},
  volume        = {10},
  number        = {2},
  pages         = {024605},
  year          = {2025},
  doi           = {10.1103/PhysRevFluids.10.024605}
}

@inproceedings{um2020solverintheloop,
  author        = {Um, Kiwon and Brand, Robert and Fei, Yun Raymond and Holl, Philipp
                   and Thuerey, Nils},
  title         = {Solver-in-the-Loop: Learning from Differentiable Physics to
                   Interact with Iterative {PDE}-Solvers},
  booktitle     = {Advances in Neural Information Processing Systems},
  volume        = {33},
  pages         = {6111--6122},
  publisher     = {Curran Associates, Inc.},
  year          = {2020}
}

@article{wang2017reynoldsdiscrepancy,
  author        = {Wang, Jian-Xun and Wu, Jin-Long and Xiao, Heng},
  title         = {Physics-Informed Machine Learning Approach for Reconstructing
                   {Reynolds} Stress Modeling Discrepancies Based on {DNS} Data},
  journal       = {Physical Review Fluids},
  volume        = {2},
  number        = {3},
  pages         = {034603},
  year          = {2017},
  doi           = {10.1103/PhysRevFluids.2.034603}
}

@article{fischer2022nekrs,
  author        = {Fischer, Paul and Kerkemeier, Stefan and Min, Misun and Lan,
                   Yu-Hsiang and Phillips, Malachi and Rathnayake, Thilina and
                   Merzari, Elia and Tomboulides, Ananias and Karakus, Ali and
                   Chalmers, Noel and Warburton, Tim},
  title         = {{NekRS}, a {GPU}-Accelerated Spectral Element {Navier--Stokes}
                   Solver},
  journal       = {Parallel Computing},
  volume        = {114},
  pages         = {102982},
  year          = {2022},
  doi           = {10.1016/j.parco.2022.102982}
}

@inproceedings{jung2025hybrid,
  author       = {Jung, Junoh and Constantinescu, Emil M.},
  title        = {Hybrid Physics--Machine Learning Framework Toward Efficient Simulation of Turbulent Flows on an Exascale Platform},
  booktitle    = {78th Annual Meeting of the APS Division of Fluid Dynamics},
  address      = {Houston, TX, USA},
  month        = nov,
  year         = {2025},
  note         = {Abstract C12.13, oral presentation},
  url          = {https://meetings-archive.aps.org/dfd/2025/c12/13/}
}

@inproceedings{jung2026hybrid,
  author        = {Jung, Junoh and Constantinescu, Emil M. and Balin, Riccardo and
                   Lusch, Bethany},
  title         = {A Hybrid Physics--Machine-Learning Framework for Enhancing
                   Coarse-Grid Spectral Element Simulations for Large-Scale Computing},
  booktitle     = {AIAA AVIATION 2026 Forum},
  publisher     = {American Institute of Aeronautics and Astronautics},
  address       = {San Diego, CA},
  year          = {2026},
  doi           = {10.2514/6.2026-4474},
  note          = {{AIAA} Paper 2026-4474}
}

@article{jung2026weakform,
  author        = {Jung, Junoh and Constantinescu, Emil M.},
  title         = {Learning Differentiable Weak-Form Corrections to Accelerate Finite Element Simulations},
  journal       = {arXiv preprint arXiv:2601.20019},
  year          = {2026},
  eprint        = {2601.20019},
  archivePrefix = {arXiv},
  primaryClass  = {cs.LG},
  doi           = {10.48550/arXiv.2601.20019}
}

@article{kang2023differentiabledg,
  author        = {Kang, Shinhoo and Constantinescu, Emil M.},
  title         = {Differentiable {DG} with Neural Operator Source Term Correction},
  journal       = {arXiv preprint arXiv:2310.18897},
  year          = {2023},
  eprint        = {2310.18897},
  archivePrefix = {arXiv},
  primaryClass  = {physics.flu-dyn},
  note          = {Revised version, 2025},
  doi           = {10.48550/arXiv.2310.18897}
}

@article{kang2023subgridnode,
  author  = {Kang, Shinhoo and Constantinescu, Emil M.},
  title   = {Learning Subgrid-Scale Models with Neural Ordinary Differential Equations},
  journal = {Computers \& Fluids},
  volume  = {261},
  pages   = {105919},
  year    = {2023},
  doi     = {10.1016/j.compfluid.2023.105919}
}

@article{rathgeber2017firedrake,
  author        = {Rathgeber, Florian and Ham, David A. and Mitchell, Lawrence and
                   Lange, Michael and Luporini, Fabio and McRae, Andrew T. T. and
                   Bercea, Gheorghe-Teodor and Markall, Graham R. and Kelly, Paul H.
                   J.},
  title         = {{Firedrake}: Automating the Finite Element Method by Composing
                   Abstractions},
  journal       = {ACM Transactions on Mathematical Software},
  volume        = {43},
  number        = {3},
  pages         = {24:1--24:27},
  year          = {2017},
  doi           = {10.1145/2998441}
}

@article{Mitusch2019DolfinAdjoint,
  author    = {Mitusch, Sebastian K. and Funke, Simon W. and
               Dokken, J{\o}rgen S.},
  title     = {{dolfin-adjoint} 2018.1: automated adjoints for
               {FEniCS} and {Firedrake}},
  journal   = {Journal of Open Source Software},
  year      = {2019},
  volume    = {4},
  number    = {38},
  pages     = {1292},
  doi       = {10.21105/joss.01292},
  url       = {https://doi.org/10.21105/joss.01292},
  publisher = {The Open Journal}
}

@article{barkley1996floquet,
  author  = {Barkley, Dwight and Henderson, Ronald D.},
  title   = {Three-Dimensional Floquet Stability Analysis of the Wake of a Circular Cylinder},
  journal = {Journal of Fluid Mechanics},
  volume  = {322},
  pages   = {215--241},
  year    = {1996},
  doi     = {10.1017/S0022112096002777}
}

@article{barwey2025meshsr,
  author        = {Barwey, Shivam and Pal, Pinaki and Patel, Saumil and Balin,
                   Riccardo and Lusch, Bethany and Vishwanath, Venkatram and Maulik,
                   Romit and Balakrishnan, Ramesh},
  title         = {Mesh-Based Super-Resolution of Fluid Flows with Multiscale Graph
                   Neural Networks},
  journal       = {Computer Methods in Applied Mechanics and Engineering},
  volume        = {443},
  pages         = {118072},
  year          = {2025},
  doi           = {10.1016/j.cma.2025.118072}
}

@article{brachet1983tgv,
  author        = {Brachet, Marc E. and Meiron, Daniel I. and Orszag, Steven A. and
                   Nickel, B. G. and Morf, Rudolf H. and Frisch, Uriel},
  title         = {Small-Scale Structure of the {Taylor--Green} Vortex},
  journal       = {Journal of Fluid Mechanics},
  volume        = {130},
  pages         = {411--452},
  year          = {1983},
  doi           = {10.1017/S0022112083001159}
}

@article{posdziech2007cylinder,
  author        = {Posdziech, Oliver and Grundmann, Roger},
  title         = {A Systematic Approach to the Numerical Calculation of Fundamental
                   Quantities of the Two-Dimensional Flow over a Circular Cylinder},
  journal       = {Journal of Fluids and Structures},
  volume        = {23},
  number        = {3},
  pages         = {479--499},
  year          = {2007},
  doi           = {10.1016/j.jfluidstructs.2006.09.004}
}

@article{vanrees2011tgv,
  author  = {{van Rees}, Wim M. and Leonard, Anthony and Pullin, D. I. and Koumoutsakos, Petros},
  title   = {A Comparison of Vortex and Pseudo-Spectral Methods for the Simulation of Periodic Vortical Flows at High {Reynolds} Numbers},
  journal = {Journal of Computational Physics},
  volume  = {230},
  number  = {8},
  pages   = {2794--2805},
  year    = {2011},
  doi     = {10.1016/j.jcp.2010.11.031}
}

@article{williamson1996vortex,
  author        = {Williamson, C. H. K.},
  title         = {Vortex Dynamics in the Cylinder Wake},
  journal       = {Annual Review of Fluid Mechanics},
  volume        = {28},
  pages         = {477--539},
  year          = {1996},
  doi           = {10.1146/annurev.fl.28.010196.002401}
}
 \begin{center}
	\scriptsize \framebox{\parbox{5in}{Government License (will be removed at publication):
			The submitted manuscript has been created by UChicago Argonne, LLC,
			Operator of Argonne National Laboratory (``Argonne").  Argonne, a
			U.S. Department of Energy Office of Science laboratory, is operated
			under Contract No. DE-AC02-06CH11357.  The U.S. Government retains for
			itself, and others acting on its behalf, a paid-up nonexclusive,
			irrevocable worldwide license in said article to reproduce, prepare
			derivative works, distribute copies to the public, and perform
			publicly and display publicly, by or on behalf of the Government. The Department of Energy will provide public access to these results of federally sponsored research in accordance with the DOE Public Access Plan. http://energy.gov/downloads/doe-public-access-plan.
}}
	\normalsize
\end{center}
\end{document}